\documentclass[twocolumn,natbib,aps,prl,amsmath,amsfonts,nofootinbib,preprintnumbers,superscriptaddress,secnumarabic]{revtex4-1}

\usepackage{physics}
\usepackage{mathtools}
\usepackage{graphicx}% Include figure files
\usepackage{bm}% bold math
\usepackage{amssymb,amsmath,latexsym,mathrsfs}
\usepackage[usenames,dvipsnames]{color}
\usepackage[breaklinks,colorlinks,urlcolor=magenta,citecolor=magenta,linkcolor=magenta]{hyperref}
\usepackage{float}
\usepackage{xcolor}
\usepackage{units}
\definecolor{linkcolor}{rgb}{0.0, 0.47, 0.75}
\definecolor{citecolor}{rgb}{1.0, 0.5, 0.0}
\hypersetup{
  linkcolor  = linkcolor,
  citecolor  = linkcolor,
  urlcolor   = linkcolor,
  colorlinks = true
}

\begin{document}

\title{Continuous spontaneous localization as the white-noise limit of spontaneous unitarity violation}

\author{Aritro Mukherjee}
\author{Jasper van Wezel}

\affiliation{Institute for Theoretical Physics Amsterdam,
University of Amsterdam, Science Park 904, 1098 XH Amsterdam, The Netherlands}

%\date{\today}
\begin{abstract}
Objective collapse theories propose modifications to Schr\"odinger's equation that solve the quantum measurement problem by interpolating between microscopic quantum dynamics and projective evolution of macroscopic objects. Colored-noise driven collapse theories extending the equilibrium description of spontaneous symmetry breaking to spontaneous violations of unitarity (SUV) in quantum dynamics were recently shown to possess a Markovian white noise limit when applied to initial two-state superpositions. Here, we show that this limit coincides with a subclass of continuous spontaneous localization (CSL) models collapsing in a basis of spatially localised energy eigenstates. We show that the energy expectation value remains conserved throughout this process, and we also extend the model to a form that can be applied to any initial state. We furthermore show that, as for the SUV models, the emergence of Born rule statistics in the Markovian limit is enforced by a fluctuation-dissipation relation which results in ensemble averaged probability densities following a linear quantum semi-group guaranteeing the absence of superluminal signaling.
\end{abstract}

\maketitle

\noindent\emph{\textbf{Context and introduction}} ---
Despite the tremendous success of quantum mechanical predictions of ensemble averaged quantities, unitary time evolution is at odds with the observation of single measurement outcomes. The resolution of this so-called quantum measurement problem remains one of the central open questions in modern physics~\cite{Bassi_03_PhyRep, leggett2005quantum, overview, arndt2014testing, carlesso2022present,Wezel10}. Objective collapse theories aim to solve it by introducing small modifications to Schr\"odinger's equation in such a way that the unitary time evolution of microscopic particles is not affected in any noticeable way, while the effect of the modifications dominate the dynamics in the macroscopic regime and cause quantum superpositions of large objects to reduce to classical states~\cite{BohmBub_66_RevModPhys, Pearle_76, Pearle_89_PRA, Gisin84, Ghirardi_1986, Diosi_87_PLA, Ghirardi_90_PRA, percival95, Penrose_96, Wezel10, aritro1,aritro2}. One particular class of recently introduced objective collapse theories, known as models of spontaneous unitarity violation (SUV)~\cite{Wezel10,Wezel_2008,aritro1,aritro2,Mertens_PRA_21,Mertens22}, is based on the observation that the unitarity of Schr\"odinger's equation is unstable to even infinitesimal perturbations in the thermodynamic limit, in the same way that equilibrium macroscopic objects may be infinitely sensitive to perturbations breaking symmetries of their governing Hamiltonian~\cite{Wezel10,Wezel_2008,Wezel_SSBlecturenotes}. 

In fact, the observation that quantum measurement may be seen as a dynamic incarnation of the process of spontaneous symmetry breaking, holds generally. This is because any macroscopic measurement machine must have a broken symmetry in order to be accessible to human observers, and post-measurement states of the macroscopic machine must be degenerate in energy and related by a symmetry operation for the machine to be unbiased. In the usual von Neumann strong measurement description of quantum measurement~\cite{Von_Neumann2018-bo}, this implies that a partly symmetric superposition is created during the first stage of measurement, while this symmetry is fully broken when a final outcome is registered. Since SUV is the natural extension of spontaneous symmetry breaking from equilibrium to dynamical conditions~\cite{Wezel10}, it naturally applies to quantum measurements. 

Moreover, unlike other objective collapse theories, the preferred basis of quantum state reduction in SUV is uniquely determined by the Hamiltonian of the measurement machine. It corresponds to the set of symmetry-broken states, which are robust to both perturbations and decoherence~\cite{Wezel_SSBlecturenotes}. Moreover, energy is naturally conserved during SUV dynamics, because all symmetry-broken states are degenerate, as opposed to the states involved in other approaches~\cite{Bassi2013Review}. For the usual case of translation symmetry being broken by a measuring device, the symmetry broken states resulting from SUV dynamics are both perfectly spatially localised and energy eigenstates in the thermodynamic limit. Notice however that spontaneous symmetry breaking actually occurs before this limit, in large but finite devices responding to finite but immeasurably small perturbations~\cite{Wezel_SSBlecturenotes}. Despite this fact, no fine-tuning or ad-hoc choice of perturbations are needed to obtain SUV dynamics, even for finite systems, because given a generic perturbation, only those components coupling to an order parameter, will have any effect within a measurable time~\cite{Wezel10,Wezel_SSBlecturenotes}. 

In this context, it was recently shown that SUV models for the quantum measurement of an object superposed over two states has a generic form containing distinct non-linear and stochastic (noisy) contributions to the unitarity-breaking perturbation~\cite{Mertens24,aritro1}. Moreover, for a wide class of colored-noise driven SUV models, it was shown that the limit in which the stochastic contribution has vanishing correlation time, the so-called white-noise limit, yields a unique Markovian form~\cite{aritro2}. 

Here, we establish the relationship between these models of spontaneous unitarity violation, and a second well-known class of objective collapse theories, the so-called continuous spontaneous localization (CSL) models~\cite{Bassi_03_PhyRep,Bassi2013Review, Ghirardi_90_PRA}. We show that CSL-type stochastic differential equations with collapse operators projecting onto (energetically degenerate) symmetry breaking states appear as an effective model, which is weakly equivalent to the colored noise driven SUV dynamics in the white-noise limit. Despite their agreement on noise-averaged state dynamics in this limit, however, we emphasize that the two models are based on distinct physical assumptions and give rise to distinct physical predictions in the presence of correlated noise.

We then continue to show that the CSL form of the white-noise limit for the two-state SUV model suggests an extension applicable to arbitrary initial wave functions, superposed over any discrete or even continuous set of initial states. This model is explicitly shown to yield Born rule probabilities for individual measurement outcomes, while the effective dynamics of an ensemble of measurements follow a linear (Markovian) master equation of the Gorini-Kossakowsky-Sudarshan-Lindblad (GKSL) form. This is enforced by a fluctuation dissipation relation between the stochastic and deterministic components of the theory which results in a stochastic Schr\"odinger equation of the CSL form, guaranteeing the impossibility of faster-than-light communication~\cite{Gisin:1989sx,Bassi2015}. 

While the model introduced here presents an extension for the white-noise limit of SUV theories, it is not straightforward to construct colored-noise models that flow to it in the limit of vanishing correlation time while obeying all requirements for objective collapse theories also at finite correlation times. We discuss some of the challenges encountered in constructing such theories.

%%%%%%%%%%%%%%%%%%%
\noindent\emph{\textbf{Spontaneous Unitarity Violation}} ---
Models for spontaneous unitarity violation are extensions of the well-established equilibrium theory of spontaneous symmetry breaking (SSB) to the dynamical realm~\cite{Wezel_SSBlecturenotes,Wezel_2008,Wezel10}. As such, they are based around a set of common principles: first, objective collapse is argued to be a dynamical form of spontaneous symmetry breaking, in which a superposition of symmetry-broken states of a macroscopic object evolve to just one such state. The single outcome breaks an internal symmetry of the object, while the irreversible dynamics resulting in that state breaks the unitarity of time evolution.

Second, this process happens ``spontaneously'' for sufficiently large objects. Here, spontaneously is used in the same way as in equilibrium SSB. That is, the susceptibility of an object to symmetry-breaking perturbations scales with the size of the system, so that an absolutely minute, immeasurably small, perturbation suffices to break the symmetry of macroscopic objects~\cite{Wezel_SSBlecturenotes}. This makes the symmetry breaking unavoidable under any practical circumstances~\cite{Wezel_2008}, but it also means that the modification of Schr\"odinger's equation in SUV models must have a physical origin (outside of standard quantum mechanics). In particular, this means the stochastic term must be physical and hence have non-zero correlation time. The white-noise limit can only ever be considered an effective description for models of SUV. 
%
%%%%%%%%%%%%%%%%%%%%%%%%%%
\begin{figure}[t]
\includegraphics[width=\columnwidth]{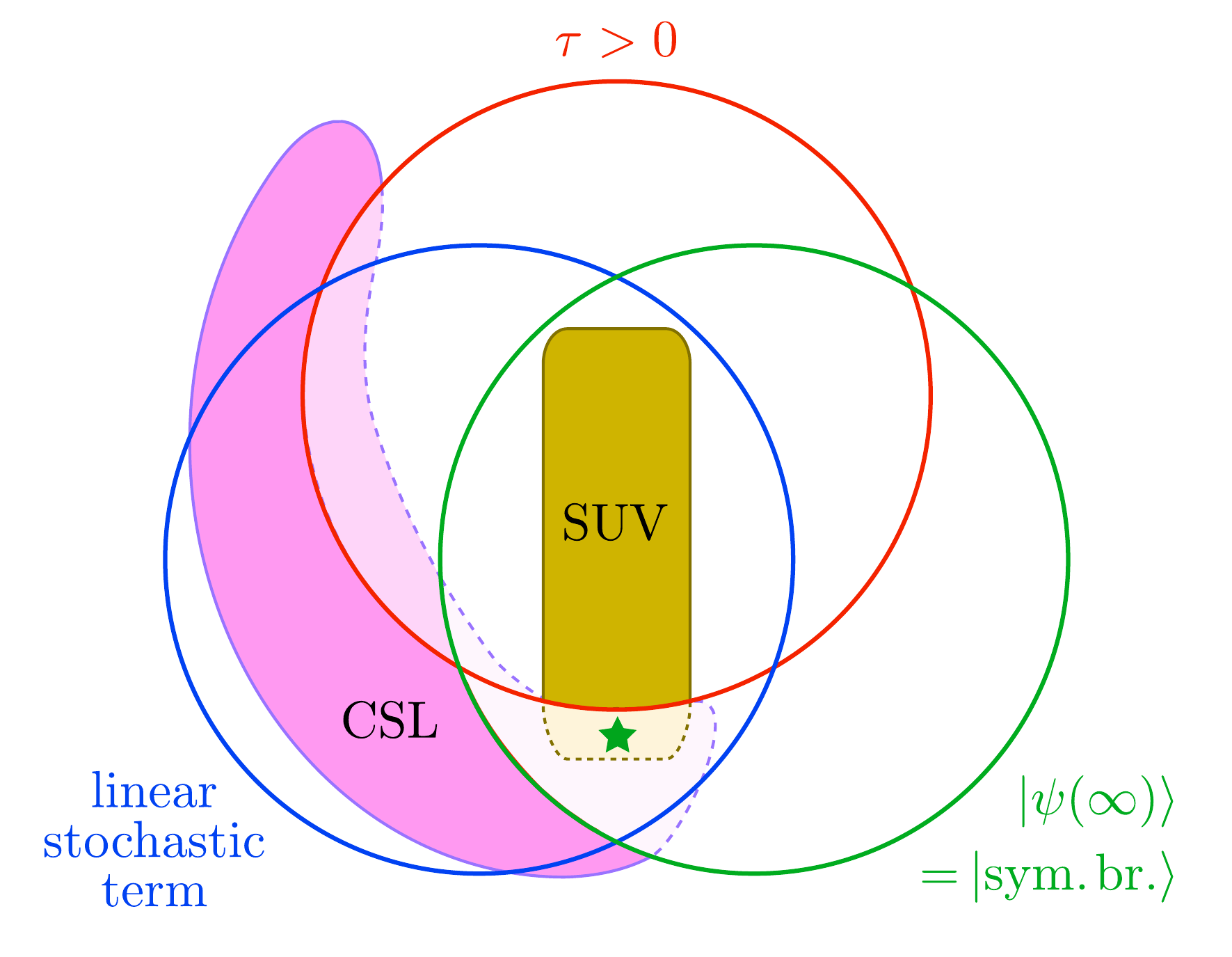}
\caption{\label{fig1} Venn diagram showing the relation between two classes of objective collapse theories. Any model for Spontaneous Unitarity Violation (SUV) falls in he central (brown) area, collapsing into a basis of symmetry-breaking states and coupling linearly to an arbitrary (possibly non-Gaussian) stochastic process. They may be extended to include uncorrelated white noise ($\tau=0$) as an idealized, mathematical limit of the physical correlated-noise scenario~\cite{Wezel10,Mertens22,aritro2}. A specific example of such a limiting model is given by Eq.~\eqref{eq:two-state} and indicated by the green star. Any model of Continuous Spontaneous Localization (CSL) on the other hand, falls into the pink shaded area on the left of the diagram. Models in this class do not necessarily possess any of the SUV constraints. Since they assume the time evolution itself to contain an inherent stochastic modification, CSL models are generally formulated in the white-noise limit. Their stochastic modification often couples linearly to the quantum state, but non-linear contributions are allowed. CSL models may be formulated with respect to any collapse basis. Common choices include the position and mass density bases~\cite{Bassi_03_PhyRep,Bassi2013Review}. Here, we show that the white-noise limit of SUV models coincides with CSL equations based on a linear coupling to a stochastic process and collapsing into a symmetry-breaking basis. For a two-state system, this particular CSL model is described by Eq.~\eqref{eq:two-state_CSL}, and since it is equivalent to Eq.~\eqref{eq:two-state}, it is also indicated by the green star.}
\end{figure}
%%%%%%%%%%%%%%%%%%%%%%%%%%

Third, for the perturbation to have any effect, it must couple to the order parameter associated with the broken internal symmetry of the macroscopic object. This requirement also guarantees that the coupling strength scales with system size, so that the perturbation has a vanishingly small effect on microscopic states. Moreover, it identifies the preferred basis of objective collapse (the basis of classical states that macroscopic objects collapse into) to necessarily be the basis of symmetry broken states~\cite{Wezel10}. This basis of states coincides with both classical states as defined in condensed matter theory~\cite{Wezel_SSBlecturenotes}, and pointer states as defined by stability against external influences~\cite{zurek_1981}. 

Finally, it is commonly assumed in SUV models that the necessary stochastic component in the modification to Schr\"odinger's equation is independent of the state to be measured and hence enters the description as a linear operator coupling to an independent noise process~\cite{Mertens22,Mertens24,aritro2}. This way, it is guaranteed that Born's rule emerges from the collapse dynamics, rather than being introduced through the equilibrium probability distribution function of the noise.

\noindent\emph{\textbf{Continuous Spontaneous Localization}} --- 
The family of CSL models was originally introduced as a continuous extension of the Ghirardi-Rimini-Weber (GRW) model, which features random and discrete localizing events interspersing the continuous Schr\"odinger evolution~\cite{Ghirardi_1986}. The localizing modification is made continuous in CSL models, resulting in a non-linear, non-unitary, stochastic equation that is usually formulated in the white-noise limit~\cite{Bassi_03_PhyRep,Bassi2013Review, Ghirardi_90_PRA, Pearle_89_PRA, Adler2007_col0,Adler_2008_col1,Bassi_2002_col4,Pearle_1996_col6,Ferialdi_2012_col7}.

As in the original GRW model, the stochastic term in CSL models is often interpreted to be a modification of the law of time evolution, causing wave functions to evolve along stochastic rather than a deterministic paths. As such, it does not necessarily stem from a physical field or influence, but is rather inherent to time evolution itself. In contrast to SUV, the white-noise form the stochastic term in CSL models is thus not commonly assumed to be a limit of any colored noise, but rather stands on it own~\cite{Bassi2013Review}. 

The relation between the stochastic and deterministic modifications of Schr\"odinger's equation in CSL models is fixed by the requirement that the noise-averaged evolution of its density matrix is described by a linear GKSL equation. This simultaneously guarantees norm conservation, adherence to Born's rule, and the absence of superluminal signaling~\cite{Gisin:1989sx,Bassi2015}. 

Finally, for CSL theories to agree with standard quantum dynamics for microscopic states, but yield objective collapse for macroscopic objects, they need to be formulated with respect to a quantity that scales with system size. There is no generic choice, and different physical observables have been proposed to fill the role~\cite{Bassi2013Review}. In contrast to SUV, this choice of observable introduces the preferred basis in which the objective collapse dynamics of CSL takes place, rather than the basis being determined by the symmetries of the Hamiltonian.

The schematic diagram in Fig.~\ref{fig1} summarizes the most prominent characteristics of SUV and CSL models, and their potential areas of overlap.

\noindent\emph{\textbf{Two state model}} ---
In the limit of vanishing correlation time for the stochastic contribution, the SUV dynamics of a two-state superposition obeying all above requirements has been shown to have the unique norm-preserving form~\cite{aritro2}: 
\begin{align}
\label{eq:two-state}
i\hbar \, d\ket{\psi} = &\hat{H}\ket{\psi}dt \notag \\
& + i J \langle\hat{S}_z\rangle \left(\hat{S}_z - \langle\hat{S}_z\rangle  \right)\ket{\psi}dt \notag \\
& +  i \sqrt{\mathcal{D}} \left(\hat{S}_z - \langle\hat{S}_z\rangle  \right)  \ket{\psi} \circ dW_t.
\end{align}
Here the wave function, $\ket{\psi}$, is time dependent and the operator $\hat{H}$ is the standard, possibly time-dependent, Hamiltonian for the combined, entangled system of the microscopic object being measured and the macroscopic measurement apparatus. The initial state is assumed to be a two-state superposition, in which the two states of the macroscopic device are distinct symmetry-breaking states corresponding to two possible measurement outcomes (for example with a pointer pointing in different directions). The operator $\hat{S}_z$ denotes the action of the order parameter in the subspace of the two symmetry breaking states.

The second term in Eq.~\eqref{eq:two-state} is a non-linear and non-unitary version of the symmetry breaking interaction appearing in the usual, equilibrium theory of spontaneous symmetry breaking~\cite{Wezel_SSBlecturenotes,Mertens22}. Here, the combination $\hat{S}_z\langle \hat{S}_z \rangle = \hat{S}_z\bra{\psi}\hat{S}_z \ket{\psi}$ should be interpreted as a non-linear operator acting on a single state, and not as an (ensemble averaged) expectation value.
Moreover, this combination is the generic form for a dissipative self-interaction. The term $\langle\hat{S}_z\rangle$ appearing in $\left(\hat{S}_z - \langle\hat{S}_z\rangle \right)$ is a purely geometric term proportional to the identity operator, which is included only to ensure norm conservation, and which does not affect any relative amplitudes~\cite{aritro2}. It may be left out altogether upon redefining the physical expectation value~\cite{Wezel10,Mertens22}. 

Notice that because $\hat{S}_z$ is an order parameter, it scales extensively with system size, $\mathcal{N}$~\cite{Wezel10,Wezel_SSBlecturenotes}: $\hat{S}_z = \mathcal{N} \hat{\sigma}_z$ with $\hat{\sigma}_z$ a Pauli matrix. To ensure that the term $\propto \langle\hat{S}_z\rangle \hat{S}_z$ remains extensive, we must demand that the coupling strength $J$ scales inversely with system size, just as in equilibrium SSB~\cite{vanwezelAmJPhys,Wezel_SSBlecturenotes}: $J=\mathcal{J}/\mathcal{N}$. Furthermore, the amplitude $\mathcal{J}$ is assumed to be immeasurably small, so that the non-unitary evolution is negligible in the quantum state evolution of microscopic objects, even though the modifications to Schr\"odinger's equation are always present. The non-commuting effects of vanishing $\mathcal{J}$ and diverging system size $\mathcal{N}$ signal the presence of an infinite susceptibility in the thermodynamic limit, in precise analogy to spontaneous symmetry breaking under equilibrium conditions~\cite{Wezel_SSBlecturenotes,Wezel_2008}. 

The third term in Eq.~\eqref{eq:two-state} is a non-unitary stochastic contribution to the order parameter dynamics. It represents coupling to an uncontrollable and unpredictable influence, giving rise to an immeasurably small diffusion constant $\mathcal{D}$. The factor $dW_t$ indicates the Gaussian increment for a standard Wiener process (i.e. uncorrelated or white noise, corresponding to the increments in Brownian motion, $W_t=\int\,dW_t$ with $W_0=0$)~\cite{revuz1999continuous,oksendal2003stochastic,gardiner2004handbook}. We use the convention that $dW_t$ is sampled from a Gaussian distribution with standard deviation $\sqrt{dt}$ and implies the standard time independent expectation values $\mathbb{E}\left[dW_t\right] = \mathbb{E}\left[W_t\right]=0$, $\mathbb{E}\left[dW_t\,dW_s\right]=0$ for $t\neq s$ and $\mathbb{E}\left[dW_t^2\right]=dt$ or simply, $dW_t^2=dt$ ~\cite{gardiner2004handbook, oksendal2003stochastic}. Here $\mathbb{E}$ indicates the average over an ensemble of Wiener process realizations. The open circle, $\circ$, preceding $dW_t$ in Eq.~\eqref{eq:two-state} indicates the Stratonovich, rather than the Ito product~\cite{aritro2}. Notice that $dW_t$ has units of square root time, and the diffusion constant $\mathcal{D}$ has units of energy squared times time. 

The stochastic influence is assumed to be physical, arising from sources beyond quantum mechanics, and not an effective description of the average influence of unobserved quantum degrees of freedom~\cite{Wezel10,aritro2}. This implies in particular that the white noise process of Eq.~\eqref{eq:two-state} should be seen as the idealized or effective white-noise limit of a physical colored-noise process with nonzero correlation time~\cite{Mertens24,aritro2}. 

\noindent\emph{\textbf{CSL from SUV}} ---
It was previously shown for the SUV model of Eq.~\eqref{eq:two-state} that if the deterministic dissipation term proportional to $\mathcal{J}$ and the fluctuation term proportional to $\sqrt{\mathcal{D}}$ obey a fluctuation-dissipation relation (FDR) of the form $\hbar\mathcal{J}=2\mathcal{D}\mathcal{N}$, the noise-averaged measurement outcomes obey Born statistics, while the noise-averaged evolution of the density matrix is described by a linear GKSL equation~\cite{aritro2}. Indeed, enforcing the FDR, Eq.~\eqref{eq:two-state} may be expressed in Ito's convention, and is then seen to coincide with a previously proposed form of CSL dynamics~\cite{Bassi_03_PhyRep,Bassi2013Review, Ghirardi_90_PRA}:
\begin{align}
\label{eq:two-state_CSL}
\, d\ket{\psi} = &-\frac{i}{\hbar} \hat{H}\ket{\psi}dt  -  \frac{\mathcal{J}\mathcal{N}}{4\hbar} \bigg(\hat{\sigma}_z - \langle\hat{\sigma}_z\rangle  \bigg)^2\ket{\psi}dt \notag \\
& +   \sqrt{\frac{\mathcal{J}\mathcal{N}}{2\hbar}} \bigg(\hat{\sigma}_z - \langle\hat{\sigma}_z\rangle  \bigg)  \ket{\psi}  dW_t.
\end{align}
Here, we made essential use of the fact that the action of the order parameter on the two states included in the SUV model of Eq.~\eqref{eq:two-state} could be written in terms of the Pauli matrix $\hat{\sigma}_z$, which squares to one.

The CSL equation here agrees with the white-noise limit of two-state SUV models described by Eq.~\eqref{eq:two-state}, which thus fall in the region of overlap in Fig.~\ref{fig1}. It should be kept in mind, however, that crucially, the SUV model derives its stochasticity from a generic model with physical, colored noise~\cite{aritro2}, which converges to Eq.~\eqref{eq:two-state_CSL} only in the idealized white-noise limit.

%%%%%%%%
\noindent\emph{\textbf{Arbitrary initial states}} ---
That the correspondence between Eqs.~\eqref{eq:two-state} and~\eqref{eq:two-state_CSL} hinges on the square of $\sigma_z$ being one, suggests a way to generalize the model to a form that allows application to initial states superposed over arbitrarily many components. To do so, we first rewrite $\hat{\sigma}_z$ in terms of projectors, which square to themselves and whose products vanish: $\hat{\sigma}_z = \hat{\mathbb{P}}_0 - \hat{\mathbb{P}}_1$. Further using the fact that $1=\hat{\mathbb{P}}_0 + \hat{\mathbb{P}}_1= \langle\hat{\mathbb{P}}_0 + \hat{\mathbb{P}}_1\rangle$, we can write $(\hat{\sigma}_z-\langle\hat{\sigma}_z\rangle)\langle\hat{\sigma}_z\rangle +1-1= 2 \sum_k (\hat{\mathbb{P}}_k-\langle\hat{\mathbb{P}}_k\rangle)\langle\hat{\mathbb{P}}_k\rangle$. Likewise, we can use the fact that the sum of two Gaussian increments is itself a Gaussian increment to introduce $dW_t \equiv (dW^0_t-dW^1_t)/\sqrt{2}$, and $d\bar{W}_t \equiv (dW^0_t+dW^1_t)/\sqrt{2}$ with $dW^0_t$ and $dW^1_t$, two independent identically distributed Gaussian increments, such that $dW_t^i dW^j_t=\delta_{ij}dt$. With this, we can write $(\hat{\sigma}_z-\langle\hat{\sigma}_z\rangle)\ket{\psi}\circ dW_t + (1-1) \ket{\psi}\circ d\bar{W}_t = \sqrt{2} \sum_k(\hat{\mathbb{P}}_k-\langle\hat{\mathbb{P}}_k\rangle)\ket{\psi} \circ dW^k_t$. The final form of the model in terms of projection operators is then:
\begin{align}
\label{eq:N-state}
i \, d\ket{\psi} = &\hat{H}\ket{\psi}dt / \hbar \notag \\
& + i 2 \mathcal{J}\,\mathcal{N} \sum_k \langle\hat{\mathbb{P}}_k\rangle \left(\hat{\mathbb{P}}_k - \langle\hat{\mathbb{P}}_k\rangle  \right)\ket{\psi} dt/\hbar \notag \\
& +  i \sqrt{\mathcal{J} \mathcal{N}} \sum_k \left(\hat{\mathbb{P}}_k - \langle\hat{\mathbb{P}}_k\rangle  \right)  \ket{\psi} \circ dW^k_t / \sqrt{\hbar}.
\end{align}
Here, we inserted the fluctuation-dissipation relation $\mathcal{D}=\hbar\mathcal{J}/(2\mathcal{N})$ in the final line~\cite{aritro2}. Using the properties of the projection operators and going to the Ito representation again yields a CSL-type stochastic Schr\"odinger equation:
\begin{align}
\label{eq:N-state_CSL}
\, d\ket{\psi} = &\frac{-i}{\hbar}\hat{H}\ket{\psi}dt  - \frac{\mathcal{J}\mathcal{N}}{2\hbar}\, \sum_k  \left(\hat{\mathbb{P}}_k - \langle\hat{\mathbb{P}}_k\rangle  \right)^2\ket{\psi} dt \notag \\
& +  \sqrt{\frac{\mathcal{J} \mathcal{N}}{\hbar}} \sum_k \left(\hat{\mathbb{P}}_k - \langle\hat{\mathbb{P}}_k\rangle  \right)  \ket{\psi} dW^k_t.
\end{align}

Extending the range of $k$ in the sums of Eq.~\eqref{eq:N-state} or~\eqref{eq:N-state_CSL} to run over an arbitrary number of states rather than just two, it can be applied to initial wave functions superposed over any (countable) number of components, and gives a generic form for objective collapse dynamics in the white-noise limit. The use of projections moreover allows a straightforward generalization to (non-relativistic) field theories, in which $\ket{\psi}$ denotes a (countable) superposition of field configurations. Equation~\eqref{eq:N-state_CSL} can thus be applied to objective collapse dynamics, but also for example to the dynamics of a symmetry-breaking quantum phase transition~\cite{WezelBerry}, in which a single, ordered field configuration arises from an initially symmetric superposition.

Note that unlike the operator $\hat{\sigma}_z$ in Eq.~\eqref{eq:two-state}, the projection operators in Eq.~\eqref{eq:N-state} cannot be directly interpreted in terms of an order parameter. To still maintain a correspondence of final states to classical configurations of a macroscopic, symmetry-breaking object, the operators $\hat{\mathbb{P}}_k$ should project onto distinct symmetry-broken states. As in the equilibrium theory of spontaneous symmetry breaking, these are determined by the unperturbed Hamiltonian and can be constructed from the group theoretical description of broken and unbroken symmetries~\cite{Wezel_SSBlecturenotes}. We will return to the connection of the arbitrary-state, white-noise model of Eq.~\eqref{eq:N-state} with possible colored-noise SUV models based on a non-Hermitian order parameter field, after first examining some of its physical properties.

%%%%%%%%%%%
\noindent\emph{\textbf{Emergent Born rule statistics and absence of superluminal signalling}} ---
To verify that the dynamics generated by Eq.~\eqref{eq:N-state} or Eq.~\eqref{eq:N-state_CSL} for arbitrary initial states gives rise to Born rule probabilities and avoids any possibility of superluminal communication, we compute its noise averaged dynamics. The dynamics of the pure state projector, $\hat{\Psi}=\ket{\psi}\bra{\psi}$, follows from substituting Eq.~\eqref{eq:N-state_CSL}, applying Ito's lemma, $d\hat{\Psi}=\ket{d\psi}\bra{\psi} + \ket{\psi}\bra{d\psi} + \ket{d\psi} \bra{d\psi}$. Averaging over the noise yields the quantum master equations~\cite{Bassi_03_PhyRep,CSL} using $\mathbb{E}[dW^k_t]=0$ (which is only true in Ito's convention). Notice that $\mathbb{E}[\hat{\Psi}] = \hat{\rho}$, where $\hat{\rho}$ is the density matrix or statistical operator describing a noise-averaged ensemble of measurements undergoing the dynamics of Eq.~\eqref{eq:N-state_CSL}. This yields a dephasing-type GKSL master equation of the form:
\begin{align}
    \hbar \frac{\partial \hat{\rho}}{\partial t} = -i \left[ \hat{H} \,,\, \hat{\rho} \right] + \mathcal{JN} \left( \sum_k \hat{\mathbb{P}}_k \hat{\rho} \hat{\mathbb{P}}_k - \hat{\rho} \right).
    \label{Eq:master1}
\end{align}
The dephasing-like dynamics of the density matrix in this expression is characterized by the decay of its off-diagonal elements, while its diagonal elements remain constant in time. An initial pure density matrix thus reduces to a mixed density matrix with only diagonal components at long times, describing an ensemble of measurements with distinct realizations of the noise, without tracing over any part of the system or its environment. The absence of a trace operation here, as opposed to the effective evolution encountered in decoherence, may be employed to construct protocols allowing SUV dynamics to be experimentally distinguished from environmentally induced dephasing in mesoscopic experiments~\cite{Wezel_2012}.

The probabilities on the diagonal of the final density matrix are the same as the squared wave function components appearing in the pure initial ensemble since these remain constant throughout the evolution. Thus, each diagonal element evolves according to a Martingale process, which ensures the emergence of Born statistics~\cite{Pearle_89_PRA,Pearle1984_bookRef,Ghirardi_90_PRA,Bassi_03_PhyRep,aritro2}. Furthermore, the linearity of Eq.~\eqref{Eq:master1} describing the time evolution of the noise-averaged density matrix, and the fact that the GKSL master equation is a linear quantum dynamical semi-group~\cite{Lindblad1976,GKS76}, guarantee the impossibility of any superluminal signaling using this dynamics~\cite{Gisin:1989sx,Bassi2015,aritroArxiv}. Both the emergence of Born statistics and the impossibility of superluminal signaling thus follow from the fluctuation-dissipation relation, which ensures that Eq.~\eqref{eq:N-state_CSL} adheres to the unique form of white-noise quantum stochastic processes yielding a linear GKSL master equation~\cite{Unique1_bassi13}.

The operators $\hat{\mathbb{P}}_k$ in Eq.~\eqref{Eq:master1} represent projections onto a set of symmetry-breaking states of the measurement device. Because these states are related by a symmetry operation, their energy expectation values are equal to one another for any system size. Moreover, these states become strictly orthogonal in the thermodynamic limit, implying that energy is then strictly conserved. This can be seen most straightforwardly by first writing the change in the average energy of an ensemble of closed systems evolving under Eq.~\eqref{Eq:master1} as:
\begin{align}
\label{eq:dEdt}
    dE/dt\, =\frac{\mathcal{JN}}{\hbar} \bigg( \sum_k\mathrm{Tr}\big[\hat{\rho}\,\hat{\mathbb{P}}_k\hat{H}\hat{\mathbb{P}}_k\big] - \mathrm{Tr}\big[\hat{\rho}\,\hat{H}\big]\bigg).
\end{align}
It is convenient to evaluate this expression using a basis of symmetric states, $\ket{n,s}$, such that $\hat{H}\ket{n,s}=E_{n,s}\ket{n,s}$. Here, the quantum number $s$ is the eigenvalue of the symmetry operator $\hat{S}$, which commutes with $\hat{H}$, and the label $n$ encodes all other quantum numbers of the system. The symmetry-broken states can then be written as $\ket{n,k}=\sum_s \psi_n(s,k) \ket{n,s}$, and $\hat{\mathbb{P}}_k = \sum_{n} \ket{n,k}\bra{n,k}$. Notice that these states are normalized, such that $\sum_s |\psi_n(s,k)|^2 = 1$ for any value of $k$ or $n$. For finite-sized systems, however, the symmetry-breaking states (with index $k$ corresponding to the value of the order parameter) in general form an overcomplete basis, so that $\sum_k |\psi_n(s,k)|^2$ becomes strictly equal to one only in the thermodynamic limit~\cite{Wezel_SSBlecturenotes}. 

Using the symmetry of the Hamiltonian, it is clear that all symmetry-related states have the same energy expectation value, and in particular that $\bra{n,k}\hat{H}\ket{n',k}=\delta_{n,n'} \tilde{E}_n$ for any value of $k$, and for any system size. This in turn implies that $\sum_k \hat{\mathbb{P}}_k \hat{H} \hat{\mathbb{P}}_k=\sum_{n,k} \ket{n,k} \bar{E}_n \bra{n,k}$. In the thermodynamic limit, this can equivalently be written as $\sum_{n,s} \ket{n,s} \bar{E}_n \bra{n,s}$, because of the orthonormality of the symmetry breaking states. Using the fact that the symmetric states $\ket{n,s}$ become degenerate in the thermodynamic limit~\cite{Wezel_SSBlecturenotes}, so that $E_{n,s}\to\bar{E}_n$, finally yields $dE/dt=0$.  

Notice that although the final steps become exact only in the thermodynamic limit, their violation in large but finite devices is in practice unobservable already for moderately-sized systems, in the same way that the spread of the wave function is practically unobservable for everyday objects~\cite{Wezel_SSBlecturenotes}.
For the generic case of measurement machines breaking translational symmetry, Eq.~\eqref{Eq:master1} implies that the final states become both precisely spatially localised, and exact energy eigenstates in the thermodynamic limit, while being localised and energy eigenstates for all practical purposes for any moderate size. 

Finally, the restriction of Eq.\eqref{Eq:master1} to initial states superposed over a countable number of components can be lifted to allow SUV dynamics starting from a continuous quantum state of the form $\ket{\psi}=\int dx\, \psi(x) \ket{x}$. To do so, the Brownian motion $dW_t^k$ with $k$ a discrete label is generalized to a Brownian sheet $dW_t^x$, with $x$ a continuous parameter, which requires a generalization of white noise to space-time white noise~\cite{WalshBook,oksendal2022spacetime} as elucidated in the supplementary material. Then, taking the continuum limit of Eq.~\eqref{eq:N-state}, as shown in detail in the supplementary material, leads to a continuum version of the dephasing GKSL master equation:
\begin{align}
    \hbar \frac{\partial \hat{\rho}}{\partial t} = -i\left[ \hat{H} \,,\, \hat{\rho} \right] + \mathcal{JN} \left( \int d\hat{\mathbb{P}}_x \, \bra{x}\hat{\rho}\ket{x} - \hat{\rho}\right).
    \label{Eq:master2}
\end{align}
Here, $\int d\hat{\mathbb{P}}_x = \int dx \, \ket{x}\bra{x}$ is the projector-valued measure used to describe spectral decompositions of unbounded operators. The continuum master equation of Eq.~\eqref{Eq:master2} has the same properties as the discrete version of Eq.~\eqref{Eq:master1}, and thus again implies the emergence of Born rule statistics and the impossibility of superluminal signaling.

%%%%%%%%%%%
\noindent\emph{\textbf{Correlated noise}} --- 
The wave function dynamics of Eq.~\eqref{eq:N-state} provides a generic model for objective collapse dynamics in the white-noise limit. Although (effectively) stochastic processes in nature may be well-approximated by white noise, however, they never have identically vanishing correlation time. Combined with the fact that all ``spontaneous'' symmetry breaking, including spontaneous unitarity violation, must be caused by a small but non-zero perturbation~\cite{Wezel_SSBlecturenotes}, this implies that 
for Eq.~\eqref{eq:N-state} to describe the white-noise limit of a physical SUV process, there must exist a colored-noise version of its dynamics. As alluded to before, the colored-noise dynamics should then also allow an interpretation in terms of an order parameter, whose form reduces to that of the projection operators in Eq.~\eqref{eq:N-state} in the white-noise limit.

In fact, a single white-noise limit may typically be reached from multiple colored-noise models~\cite{aritro2}. It is therefore not possible to uniquely associate Eq.~\eqref{eq:N-state} with a particular SUV model containing correlated noise. Given a colored-noise model, however, we can use the multi-scale noise homogenization method described in Ref.~\cite{aritro2} to obtain its effective dynamics in the limit of vanishing correlation time, and thus look for models with the correct white-noise limit. 

In particular, as explained in more detail in the supplementary material, in the context of SUV models the generic effect of noise homogenization, if it is possible, is to reduce a colored-noise driven non-linear temporal integral to a white-noise driven stochastic process, such that the two converge in probability in the limit of vanishing correlation time. The fluctuating term in Eq.~\eqref{eq:N-state} may thus be interpreted to arise from a colored-noise driven non-linear integral, whose white-noise limit follows from the multi-scale noise homogenization prescription:
\begin{align}
\label{eq:convergence}
\lim_{\tau \to 0}\, &  G\int \left(\hat{\mathbb{P}}_k-\langle\hat{\mathbb{P}}_k\rangle\right) \ket{\psi} \xi^k_t dt \nonumber\\ &\quad \longrightarrow  \sqrt{\mathcal{D}} \int \left(\hat{\mathbb{P}}_k-\langle\hat{\mathbb{P}}_k\rangle\right) \ket{\psi} \circ dW^k_t.
\end{align}
Here, $\xi^k_t$ are independent (colored) stochastic processes labelled by the index $k$, each of which evolves stochastically according to $d\xi^k=-\xi^k{dt}/{\tau} + g(\xi^k) dW^k_t$. This prescription holds if the correlated noise dynamics does not depend on the quantum state, while the quantum state itself depends linearly on the noise, and the noise with correlation time $\tau$ reaches a a steady state distribution at long times. The diffusion parameter $\mathcal{D}\propto G^2\tau$ is an effective coupling constant of the stochastic term in the limit of vanishing correlation time, which depends on $G$ and $\tau$ in such as way that $\mathcal{D}$ remains finite as $\tau$ is taken to zero.

Notice that the convergence of colored noise dynamics to the white noise evolution as the correlation time $\tau$ is taken to zero, is an instance of weak equivalence. This implies that the individual trajectories occurring in the colored-noise and white-noise models may differ, but that they agree on predictions of all probabilistic quantities~\cite{aritro2,Pavliotis2008,HorsthemkeBook2006,BoninTraversaSmallTau}. Besides the fact that the multi-scale noise homogenization explicitly yields the prescription of Eq.~\eqref{eq:convergence}, this form may also be expected based on the Wong-Zakai theorems in the regime where the $\xi_k$ dynamics is much faster than all other time scales in the problem~\cite{WongZakai1965relation,wongZakai1965convergence,WongZakai1969,WongZakaiReview, Bo_whirlProd_2013}. 

Based on Eq.~\eqref{eq:convergence}, we can consider a generic model driven by colored-noise, evolving states according to the following law:
\begin{align}
\label{eq:N-state-color}
i\hbar \, &d\ket{\psi} =  \hat{H}\ket{\psi}dt \notag \\ & + i \mathcal{N}\sum_k \left(\hat{\mathbb{P}}_k - \langle\hat{\mathbb{P}}_k\rangle  \right)\bigg[2 \mathcal{J}\langle\hat{\mathbb{P}}_k\rangle + G\,\xi^k_t \bigg] \ket{\psi} dt. 
\end{align}
Although with any noise process $\xi^k_t$, that allows the prescription of Eq.~\eqref{eq:convergence} for the limit of vanishing correlation time, this expression indeed reduces to Eq.~\eqref{eq:N-state}, it is not obvious which type of noise process one could choose for it to act as a colored-noise SUV model by itself. Choices like the Gaussian Ornstein-Uhlenbeck process or the (non-Gaussian) Brownian motion on the sphere, for which multi-scale noise homogenization is worked out in the supplemental material, allow the prescription of Eq.~\eqref{eq:convergence}, but they do not result in Born rule statistics and signaling-free dynamics starting from a generic $N$-state superposition and noise with nonzero correlation time~\cite{aritroArxiv}. We therefore leave it as a question for future research whether any particular correlated noise process $\xi^k_t$ exists, such that an expression of the form of Eq.~\eqref{eq:N-state-color} meets all requirements for a theory of spontaneous unitarity violation. 

%%%%%%%%%
\noindent\emph{\textbf{Conclusions}} ---
We showed that the model for spontaneous unitarity violation (SUV) that was recently proposed for two-state superpositions, obtains the form of a CSL-type stochastic Schr\"odinger equation in the white-noise limit. The two approaches to modeling objective quantum measurement are thus shown to overlap and agree in that limit, given a particular fluctuation-dissipation relation. This ensures that the dynamics in the white-noise limit is of the unique form yielding a dephasing-type GKSL master equation for the evolution of the noise-averaged density matrix. It is shown that, as a consequence, Born rule statistics for outcomes of quantum measurements emerge, and the impossibility of superluminal communication is guaranteed. 

We furthermore showed that the two-state dynamics in the white-noise limit can be generalized to arbitrary initial states, which again yield Born rule statistics and disallow superluminal communication. The extension of the arbitrary-state dynamics from the white-noise limit to colored-noise SUV models was shown not to be straightforward, and is left as a topic for further research.

Notice that despite their overlap in limiting equations, the SUV models are distinct from CSL and other objective collapse models. First of all, the white-noise limit in SUV models can only be considered as the idealized, mathematical limit of a temporally correlated (and possibly non-Gaussian) stochastic process. Spontaneous symmetry breaking, and therefore also spontaneous unitarity violation, requires a small but non-zero physical perturbation. For SUV in particular, this implies the stochastic contribution to the dynamics must be physical, and hence have non-zero correlation time. 

Furthermore, the basis into which states collapse during SUV dynamics is not imposed \emph{a priori}, but rather emerges in the same way that a symmetry-broken basis emerges in equilibrium spontaneous symmetry breaking. Macroscopically large objects have a diverging susceptibility to symmetry-breaking perturbations, but not to any other types. A fully generic perturbation can therefore be seen as the sum of many components that have little effect, and a single symmetry-breaking component that dominates the non-unitary dynamics and is included in the SUV model. In other words, a multitude of projections acting as collapse operators could be included in an evolution such as Eq.~\eqref{Eq:master1}, but in the limit of weak coupling strength $\mathcal{J}$, only projectors onto states breaking a symmetry of the Hamiltonian will have an observable effect in finite time. This way, SUV models avoid the preferred basis problem of alternative collapse models and ensures energy conserving dynamics.

These properties of SUV theories, and their connection to CSL models in the white-noise limit, open the way to investigating their role in the dynamics of phase transitions, as well as equilibration and thermalization, all of which require irreversible and stochastic dynamics in the quantum-classical crossover regime leading to the emergence of classical configurations.

%%%%%%%%%%
~\\
\emph{\textbf{Acknowledgement}} --- 
A.M. acknowledges supporting conversations with, A. Bassi, D. Snoke, L. Di\'osi and S.M.P. Devi.

%%%%%%%%%%%%%%%
%

%%%%%%%%%%%%%%%%%%%%%%%%%%%%%%%%

\newpage

\onecolumngrid
\newpage

\begin{center}
  \Large{Supplementary material for:\\Continuous spontaneous localization from the white-noise limit of spontaneous unitarity violation}
\end{center}

~\\~\\
{\bf This supplementary material provides details and background information supporting the calculations and arguments of the main text.}

~\\~\\

\section{Multi-scale noise homogenization}
This section introduces the application of multi-scale noise homogenization to colored noise driven dynamics in a Hilbert space. We explicitly perform this computation to obtain Eq.~(3) in the main text from Eq.~(8) for two colored noise processes, while proving the prescription of Eq.~(7) for colored noise processes possessing a well defined vanishing correlation time or white noise limit.

\subsection{Stochastic Differential Equations}
Here, we first give a brief overview of stochastic processes in general, to establish notation. The next subsection employs this to give a more detailed discussion of two specific types of stochastic processes referred to in the main text. More mathematically complete treatments are available in Refs.~\cite{SMoksendal2003stochastic,SMrevuz1999continuous,SMPavliotis2008,SMHorsthemkeBook2006}, while more physically motivated accounts may be found in Refs.~\cite{SMBreur_Petr02,SMgardiner2004handbook,SMRisken1996}. For completeness, this section reproduces parts of the discussion in the appendices of Ref.~\cite{SMaritro1,SMaritro2}.

We employ It\^o's calculus for continuous-time stochastic processes, assuming an underlying filtered probability space $(\Omega,\mathcal{F},\mathcal{F}_t,\mathbb{Q})$ endowed with a natural filtration. Here, $\Omega$ denotes the sample space encompassing all conceivable events of the stochastic process. $\mathcal{F}$, the filtration, signifies the space encompassing all event collections (i.e., all conceivable subsets of $\Omega$), and $\mathbb{Q}$ represents the probability measure associating probabilities with abstract events. The overall probability complies with $\mathbb{Q}[\mathcal{F}]=1$. The space $\mathcal{F}$, constitutes a $\sigma$-algebra. The (natural) filtration, $\{\mathcal{F}_{0\leq t<\infty}\}\subset\mathcal{F}$, comprises of a sequence of sub-$\sigma$-algebras with a causal ordering. It is dictated by all possible histories of a process leading up to time $t$. This sequence adheres to the causality condition $\mathcal{F}_{0}\subset\mathcal{F}_{t_1}\subset\mathcal{F}_{t_2}..\subset\mathcal{F}_{t}\subset\mathcal{F}$ for $0<t_1<t_2..<t$, where each $\mathcal{F}_s$ represents the set of all event collections forming a history leading up to time $0<s<t$. The process $Z_t$ ($t>0$) is termed an adapted process if $Z_t$ is $\mathcal{F}_t$ measurable, indicating that $Z_t$ possesses a well-defined probability for all possible histories leading up to time $t$, without requiring knowledge of the future (non-anticipating and thus, Markovian). All stochasitc processes considered below are assumed to be adapted processes.

An It$\mathrm{\hat{o}}$ process $Z_t$ is an adapted stochastic process, expressible either as a stochastic integral or a stochastic differential equation (SDE):
\begin{align}
Z_t &= Z_0 + \int^{t}_0A(Z_s,s)ds + \int^{t}_0 B(Z_s,s) dW_s \notag \\
\Leftrightarrow ~ dZ_t &= A(Z_t,t)dt +B(Z_t,t) dW_t.
\label{Eq:App_Ito}
\end{align}
In these expressions, the drift term $A(Z_t,t)$ and the diffusion term $ B(Z_t,t)$ are assumed to be smooth functions and $\mathcal{F}_t$ measurable for all $t$ (for a more rigorous discussion, see Refs.~\cite{SMoksendal2003stochastic,SMPavliotis2008}). The symbol $W_t=\int^t_0dW_t$ represents the standard Wiener (Brownian) process characterized by independent increments and (almost surely) continuous paths, defined with $W_0=0$ and $dW_t:=W_{t+dt} - W_{t}\sim \mathbb{N}(0, \sqrt{dt})$, which is a Gaussian distribution with mean zero and standard deviation $\sqrt{dt}$. In Hida's characterization of the Weiner process~\cite{SMHida1980}, Brownain motion may be interpreted as the temporal integral of an ideal Gaussian white noise process (random hits in time) denoted by $\eta_t\sim\,\mathbb{N}(0,1)$, i.e. $dW_t=\eta_t\,dt$ and $W_t=\int^t\eta_s\,ds$. This construction is detailed in further sections to generalize to a space-time stochastic process. Note that the Weiner process has mean zero, $\mathbb{E}_\mathbb{Q}[W_t]=0$, and variation $\mathbb{E}_\mathbb{Q}[W^2_t]=t$. Additionally, we have $\mathbb{E}_\mathbb{Q}[dW_t]=0$ and $\mathbb{E}_\mathbb{Q}[dW_tdW_s]=0$ for $t\neq s$, alongside $\mathbb{E}_\mathbb{Q}[dW_t^2]=dt$. These properties yield the `informal' It\^o multiplication rules $dt^2=dW_t dt=0$ and $dW_t^2=dt$. Importantly, smooth functions of It\^o processes remain It\^o processes themselves. 

Generalising, an $N$-dimensional, vector-valued It\^o process $\mathbf{Z}_t$, with components $\{Z^k_t\},\,k\in[1,N]$ can be defined to follow the SDE:
\begin{align}
d\mathbf{Z}_t = \mathbf{A}(\mathbf{Z}_t)dt + \mathbf{B}(\mathbf{Z}_t) d\mathbf{W}_t.
\label{Eq:Z_vec_Ito}    
\end{align}
Here $\mathbf{A}(\mathbf{Z}_t): \mathbf{R}^N\rightarrow\mathbf{R}^N$ is vector valued, being just an $N$ dimensional array, $\mathbf{B}(\mathbf{Z}_t)_{N\,\times\,M}:\mathbf{R}^N\rightarrow\mathbf{R}^{N\,\times\,M}$ is a $N\,\times\,M$ matrix valued function of $\mathbf{Z}_t$, and $d\mathbf{W}_t$ are $M$-dimensional Brownian increments with components $\{dW^k_t\}$ and $k\in[1,M]$.

The evolution of probabilities of functions of stochastic processes is studied by their transition probability density $\mathbb{T}(\mathbf{Z}_f,t_f|\mathbf{Z_0},t_0)$, which is defined as the probability of progression from the initial Dirac-delta distribution for the state of the stochastic process localized at $\mathbf{Z_0}$, to a terminal state $\mathbf{Z_f}$. The transition probabilities can be interpreted either to evolve forward in time from specified initial conditions following the Fokker-Planck-Kolmogorov (FPK) equations, or backward in time from specified terminal conditions following the Kolmogorov Backward (KB) equations. Both equations can be expressed in terms of an infinitesimal generator $\Lambda$ (also known as a characteristic operator), where $\Lambda(\mathbf{Z},t)$ evolves the transition probability density in the backward direction, while its adjoint $\Lambda^{*}(\mathbf{Z},t)$ propels it forward:
\begin{align}
\text{KB:}~-\frac{\partial}{\partial t}\mathbb{T}(\mathbf{Z}_f,t_f|\mathbf{Z},t) &= \Lambda(\mathbf{Z},t)\mathbb{T}(\mathbf{Z}_f,t_f|\mathbf{Z},t) \notag \\
\text{FPK:}~~\,\,\frac{\partial}{\partial t}\mathbb{T}(\mathbf{Z},t|\mathbf{Z}_0,t_0) &= \Lambda^{*}(\mathbf{Z},t)\mathbb{T}(\mathbf{Z},t|\mathbf{Z}_0,t_0).
\label{eqApp:FP}
\end{align}
Integrating out the terminal (initial) conditions from the Kolmogorov Backward (Fokker-Planck-Kolmogorov) equations yields the one-time probability density, $\rho(\mathbf{Z},t)$, employing $\int \mathbb{T}(\mathbf{Z}_T,t_T|\mathbf{Z},t)\rho(\mathbf{Z}_T,t_T)d\mathbf{Z}_T = \rho(\mathbf{Z},t)$ or $\int \mathbb{T}(\mathbf{Z},t|\mathbf{Z}_0,t_0)\rho(\mathbf{Z}_0,t_0)d\mathbf{Z}_0 = \rho(\mathbf{Z},t)$. This one-time probability density is also a solution of Eq.\eqref{eqApp:FP}. Explicit expressions in operator form for the generator $\Lambda(z,t)$ and its adjoint for the one dimensional It\^o process $Z_t$ are:
\begin{align}
\Lambda(Z,t) \,\,\bullet&= A(Z,t)\frac{\partial\,\,\bullet}{\partial Z} + \frac{B^2(Z,t)}{2}\frac{\partial^2\,\,\bullet}{\partial Z^2 } \notag \\
\Lambda^{*}(Z,t)\,\,\bullet &= -\frac{\partial}{\partial Z} A(Z,t)\bullet + \frac{\partial^2}{\partial Z^2 }\frac{B^2(Z,t)}{2}\,\bullet.
\label{Eq.App.FPKOp}
\end{align}
The derivation of these forms involves an integration by parts to derive one from the other, assuming the associated surface terms vanish under compact support.
The generator and its adjoint for the vector process $\mathbf{Z}_t$ are given by:
\begin{align}
\Lambda(\mathbf{Z},t)\,\,\bullet &:= \sum_i\,A_i(\mathbf{Z}) \frac{\partial\,\,\bullet}{\partial Z^i} +\,\frac{1}{2}\sum_{ij} \left[\mathbf{B}\mathbf{B}^T\right]_{ij} \,\frac{\partial^2\,\,\bullet}{\partial\,Z^i \partial Z^j}\notag\\
\Lambda^{*}(\mathbf{Z},t)\,\,\bullet& :=- \sum_i\, \frac{\partial}{\partial Z^i}\,A_i(\mathbf{Z}) \bullet+\,\frac{1}{2}\sum_{ij} \,\frac{\partial^2}{\partial\,Z^i \partial Z^j}\left[\mathbf{B}\mathbf{B}^T\right]_{ij} \,\bullet.
\end{align}
Besides the It\^o interpretation, stochastic processes with multiplicative noise can also be described using the Stratonovich interpretation, in which the term $B(Z_t)dW_t$ is replaced by $B(Z_t)\circ\,dW_t$, with $\circ$ denoting a Stratonovich product. A Stratonovich process may be defined to follow the Statonovich SDE (as opposed to It\^o SDE):
$$
d\mathbf{Z}_t = \bar{\mathbf{A}}(\mathbf{Z}_t)dt + \mathbf{B}(\mathbf{Z}_t)\circ d\mathbf{W}_t.
$$
Converting this expression to It\^o's interpretation (notation) yields:
\begin{align*}
d\mathbf{Z}_t &= \mathbf{A}(\mathbf{Z}_t)dt + \mathbf{B}(\mathbf{Z}_t) d\mathbf{W}_t \notag \\
\text{with:} ~~~
A_i(\mathbf{Z}_t) &= \bar{A}_i(\mathbf{Z}_t) + \frac{1}{2} \sum^{m}_{j=1}\sum^{n}_{k=1} \frac{\partial \,B_{ij}}{\partial \,Z_k}(\mathbf{Z}_t)\,B_{kj}(\mathbf{Z}_t).
\end{align*}

\subsection{Coloured noise processes}
To contrast time-correlated stochastic processes with (Gaussian) white noise processes that are uncorrelated in time, we refer to the former as colored noise processes. In Ref.~\cite{SMaritro2}, colored noise processes with a steady-state distribution were suggested as noise term in models of spontaneous unitarity violation. Here, we therefore focus on the Ornstein-Uhlenbeck (OU) and spherical Brownian motion (SBM) processes. The OU process is described by:~\cite{SMOU_OG1930,SMDoob42,SMgardiner2004handbook,SMRisken1996}

\begin{align}
d\xi_t = - \xi_t \,\frac{dt}{\tau} + \sqrt{\frac{2}{\tau}} dW_t
\label{Eq7OU}
\end{align}

Correlations in the OU process decay exponentially, with correlation time $\tau$. Beyond that, the process converges to a distinct steady-state Gaussian distribution centered at $\xi=0$, and given by $\rho^{\infty}(\xi)=\frac{1}{\sqrt{2\pi}}e^{\frac{-\xi^2}{2}}$ independent of initial conditions~\cite{SMgardiner2004handbook}. The analytical solution for individual noise realizations with a sharp initial condition $\xi_0$ is given by~\cite{SMgardiner2004handbook}:

\begin{align}
\xi_t = \xi_0 e^{-t/\tau} +  \sqrt{\frac{2}{\tau}} e^{-t/\tau} \int_0^t e^{s/\tau}dW_s
\label{Eq:OU_gen_sol_main}
\end{align}
Here, the Wiener measure ($dW_s$) is used to integrates a time-dependent function, giving rise to a standard It\^o integral. From Eq.~\eqref{Eq:OU_gen_sol_main}, the mean value and autocorrelation can be derived given any initial conditions. Sampling $\xi_0$ from the steady-state distribution, we find $\mathbb{E}_{\xi}[\xi_0]=0$, where the expectation value is with respect to the noise process $\xi_t$. Similarly, we find:
\begin{align}
    \mathbb{E}_{\xi}[\xi_t]   &=0 \notag \\
    \mathbb{C}[\xi_t,\xi_s]  &= e^{- |t-s|/\tau}
\end{align}
The auto-correlation function $\mathbb{C}[\xi_t,\xi_s]$ is given by $\mathbb{E}_{\xi}[\xi_t \,\xi_s] -\mathbb{E}_{\xi}[\xi_t] \mathbb{E}_{\xi}[\xi_s]$. For the conditional expectation given a sharp initial condition of $\xi_0$, one also obtains $\mathbb{E}_{\xi}[\xi_t|\xi_0]=\xi_0 \,\exp[-\frac{t-t_0}{\tau}]$, showing that on time scales larger than $\tau$, all the temporal correlations are lost. These results may alternatively be derived using the FPK equations for the OU process~\cite{SMgardiner2004handbook,SMRisken1996,SMoksendal2003stochastic}. 

The second colored noise process with non-zero correlation time and steady state distribution we consider stems from Brownian motion on a (unit) spherical manifold. It can be defined in terms of its generator, the Laplace-Beltrami operator on a sphere~\cite{SMoksendal2003stochastic}:
\begin{align}
    \Lambda(\theta,\phi)  = \frac{1}{\sin\theta}\frac{\partial}{\partial \theta}\bigg(\sin\theta\frac{\partial}{\partial\theta}\bigg) + \frac{1}{\sin^2\theta}\frac{\partial^2}{\partial\phi^2}
    \label{Eq:Laplace Belt}
\end{align}
The KB equation for the stochastic process on the surface of a sphere can then be formulated using the transition probability density $\mathbb{T}(\theta,\phi,t|\theta_0,\phi_0,t_0)$ using Eq.~\eqref{eqApp:FP} and Eq.~\eqref{Eq.App.FPKOp}:
\begin{align}
    \frac{\partial}{\partial t}\mathbb{T}(\theta_T,\phi_T,T|\theta,\phi,t)=\frac{-1}{2\tau}\Lambda(\theta,\phi)\mathbb{T}(\theta_T,\phi_T,T|\theta,\phi,t) \notag
\end{align}
From here, the terminal condition can be integrated out to obtain the one-point probability density, using the definition $\rho(\theta,\phi,t) = \int\mathbb{T} (\theta_T,\phi_T,T|\theta,\phi,t) \rho(\theta_T,\phi_T)d\theta_T d\phi_T$:
\begin{align}
\frac{\partial}{\partial t}\rho(\theta,\phi,t)=\frac{-1}{2\tau}\left[\cot\theta\frac{\partial}{\partial \theta}+\frac{\partial^2}{\partial\theta^2} \right.  +\left. \,\frac{1}{\sin^2\theta}\frac{\partial^2}{\partial\phi^2}\right]\rho(\theta,\phi,t)
\end{align}

The one-point probability density implies two It\^o processes: one for $\theta_t\in[0,\pi]$ and one for $\phi_t\in[0,2\pi]$, representing the coordinates of the Brownian walker on the sphere. The corresponding stochastic differential equations are:
\begin{align}
    d\theta_t &= \frac{1}{2\tau} \cot(\theta_t) \,dt + \frac{1}{\sqrt{\tau}}dW^{\theta}_t
    \label{Eq:theta_t}\\ 
    d\phi_t &=  \frac{1}{\sqrt{\tau}\sin(\theta_t)}dW^{\phi}_t 
    \label{Eq:phi_t}
\end{align}
From these, it is clear that the $\theta_t$ process is independent of the $\phi_t$ process at all times, whereas the evolution of $\phi_t$ depends on the instantaneous value of $\theta_t$. Because of this, the models of spontaneous unitarity violation discussed in Ref.~\cite{SMaritro2} employed the cosine of the latitude, $\xi_t=\cos(\theta_t)$, rather than the position on the sphere as the colored noise process driving state evolution. This choice provides a stochastic variable bounded between $[-1,1]$. Additionally, the steady-state distribution of the Brownian motion uniformly covers the sphere, resulting in a uniform steady-state distribution of $\xi_t$ across the range $[-1,1]$. These properties of spherical Brownian motion (SBM) have been previously shown to facilitate the emergence of Born's rule in the limit of infinite correlation time or time-independent noise~\cite{SMMertens_PRA_21,SMMertens22,SMaritro2}. 

From Eq.\eqref{Eq:theta_t}, the stochastic dynamics of the spherical Brownian motion (SBM) process is described by:
\begin{align}
d\xi_t = - \xi_t \,\frac{dt}{\tau} + \sqrt{\frac{1-\xi_t^2}{\tau}} dW_t
\label{Eq:SBM SDE}
\end{align}
The final term in this expression indicates that SBM is a multiplicative noise process, interpreted in the It\^o sense. This SBM process falls under the category of Pearson Diffusions and is also a Jacobi diffusion process (see Refs.~\cite{SMCNOSPearsonDiffPaper2008,SMCNOStimelocalPearsontoJacobiAscione2021,SMCNOSJacobiDEMNI2009518} and references therein for further details). The timescale $\tau$ corresponds to the correlation time, while the mean-reverting drift term ensures a steady-state distribution.

The mean and autocorrelation for the SBM process are obtained by considering its infinitesimal (adjoint) generator and the corresponding Fokker-Planck-Kolmogorov (FPK) equations~\cite{SMRisken1996,SMCNOSPearsonDiffPaper2008,SMCNOStimelocalPearsontoJacobiAscione2021,SMCNOSWong1964}:
\begin{align}
 \partial_t \mathbb{T}\left(\xi, t|\xi_0, t_0\right) &=  \frac{1}{\tau} \partial_\xi \left[\xi \, \mathbb{T}\left(\xi,t|\xi_0, t_0\right)\right]  
 + \frac{1}{2 \tau} \partial^2_\xi \left[ (1-\xi^2) \mathbb{T}\left(\xi, t|\xi_0, t_0\right)\right]
\end{align}
This FPK equation possesses an exact formal solution in the form of an infinite sum~\cite{SMCNOSWong1964}:
\begin{align}
\mathbb{T}\left(\xi, t|\xi_0, t_0\right) = \frac{1}{2} \sum^{\infty}_{n=0} (2 n+1) (n!)^2 P_n(\xi) P_n(\xi_0)  \exp \left[\frac{-n (n+1) |t-t_0|}{2 \tau }\right]
\label{Eq:CNOS sol}
\end{align}
Here, $P_n$ are the Legendre polynomials, given by the Rodrigues formula $P_n(x) = \frac{1}{n!\,\,2^n} \, \, \frac{d^n}{dx^n} \left(x^2-1\right)^n$. In the long-time limit ($t-t_0\to\infty$), the SBM process approaches a steady-state probability density $\rho^{\infty}(\xi)$ independent of the initial state. By taking the limit $t_0\rightarrow-\infty$ in Eq.~\eqref{Eq:CNOS sol}, only the $n=0$ term survives, leading to a completely uniform steady-state distribution, i.e., $\rho^{\infty}(\xi)=\lim_{t_0\rightarrow-\infty}\mathbb{T}\left(\xi, t|\xi_0, t_0\right)=1/2$ for $\xi\in[-1,1]$. Similarly, the unconditioned mean $\mathbb{E}_\xi[\xi_t]$ and the expected value $\mathbb{E}_\xi[\xi_t|\xi_0]$ conditioned on sharp initial conditions, $\xi_t=\xi_0$ at $t=t_0$, are found via direct integration of the transition probability density:
\begin{align*}
\mathbb{E}_\xi[\xi_t] &=\int^1_{-1}\int^1_{-1} \xi \, \mathbb{T}\left(\xi, t|\xi_0, t_0\right) \rho^{\infty}(\xi_0) \, d\xi_0 d\xi=0 \notag \\
\mathbb{E}_\xi[\xi|\xi_0] &= \int^1_{-1} \xi \, \mathbb{T}\left(\xi, t|\xi_0, t_0\right) d\xi = \xi_0 \exp\left[-\frac{t-t_0}{\tau}\right]\\
\mathbb{C}[\xi_t,\xi_s] = &\int^1_{-1}\int^1_{-1} \xi_t \,\xi_s \, \mathbb{T}\left(\xi_t, t|\xi_s, s\right) \rho^{\infty}(\xi_s)\, d\xi_t d\xi_s = \frac{1}{3} \exp[-\frac{t-t_s}{\tau}]
\end{align*}
Here, again the auto-correlation function is defined as $\mathbb{C}[\xi_t,\xi_s]=\mathbb{E}_{\xi}[\xi_t \,\xi_s] -\mathbb{E}_{\xi}[\xi_t] \mathbb{E}_{\xi}[\xi_s]$. These results reveal that, similar to the OU process, the SBM process features exponentially decaying correlations and a unique steady state distribution, albeit non-Gaussian.

\subsection{Multi-scale Noise Homogenization}
In this subsection, we discuss the analysis for obtaining Eq.~(3) from Eq.~(8) in the main text, while proving the prescription of Eq.~(7) for a class of colored noise processes possessing a well defined vanishing correlation time limit ($\tau \to 0$). A controlled perturbative method to obtain coarse-grained or time-renormalized dynamics of stochastic systems with separation of time scales has been employed in Ref.~\cite{SMaritro2} to study dynamical quantum state reduction. It is based on the so called multi-scale noise homogenization techniques developed in Refs.~\cite{SMPavliotis2008,SMHorsthemkeBook2006,SMBoninTraversaSmallTau,SMaritro2}. The calculations in this section review and generalize this technique to higher dimensions. We begin with the model introduced in the main text, given by ($\mathcal{N}$ is absorbed in the coefficients $J, G$ and we set $\hbar=1$):
\begin{align}
d\ket{\psi} &= \sum_k\, \bigg(\hat{\mathbb{P}}_k -\langle\hat{\mathbb{P}}_k\rangle\bigg)\bigg[ 2J \langle\hat{\mathbb{P}}_k\rangle+\,G\,\xi^k_t\bigg]\ket{\psi}  dt  
\notag \\
d\xi^k_t&=-\xi^k_t\,\frac{dt}{\tau} + \, g(\xi^k_t)dW^k_t
\label{Eq:App_Pair}
\end{align}
Here, the time-dependent wave function is $N$-dimensional and of the form $\ket{\psi_t}:=\ket{\psi(t)}=\sum_i\psi_i(t)\ket{i}$. The standard Hamiltonian does not admit attractive or repulsive fixed points in the dynamics, and can be neglected entirely in cases that the modifications to Schr\"odinger's equation are much larger than or commute with the Hamiltonian~\cite{SMaritro2}. We will derive the dynamics for the squared amplitudes $z_i := |\bra{i}\psi(t)\rangle|^2$. For notational convenience, we have dropped the explicit time-dependence of $\ket{\psi}$, $\psi_i$ and $z_i$, except in cases where it is required for clarity. From Eq.~\eqref{Eq:App_Pair}, the dynamical equations for $\psi_i$ are given by:
\begin{align}
d\psi_i \ket{i}\,=\left[2J \left(z_i - \sum_k\,z_k^2\right) + G\left(\xi^i_t - \sum_k \,\xi^k_t\,z_k\right)\right]\psi_i \ket{i}dt   
\end{align}

As explained in Ref.~\cite{SMaritro2}, for an arbitrary smooth function $f(\mathbf{\psi})$, the temporal integral of the form $\int f(\mathbf{\psi})\xi^k_t\,dt$ is a regular Riemann integral, since the $d\psi_k$ process has a vanishing quadratic variation ($d\psi_k^2=0$). This is because the noise $\xi_k$ is continuous while its integral is continuous and once-differentiable. Thus, there is no difference between the It\^o and Stratonovich descriptions in this case~\cite{SMHanggi94,SMIto_strat1,SMaritro2}, and using the regular rules of calculus we obtain the evolution equations for $dz_i$:
\begin{align}
    dz_i&=J_i\,dt + \,\sum_j\, G_{ij}\xi_j\,dt
\end{align}
Here, we defined $J_i = 4J z_i\left( z_i-\sum_j z_j^2\right)$ and $G_{ij} = 2G z_i\left(\delta_{ij}-z_j\right)$. To perform the multi-scale noise homogenization, the first step is to isolate a factor of $\sqrt{\tau}$ from  $G_{ij}$ by writing  $G_{ij}\to \sqrt{\tau}G_{ij}$ and $dt\to dt/\sqrt{\tau}$. This effectively coarse-grains the dynamics, since the SDE is formally understood as a discrete sum over temporal partitions, which increase as $\frac{dt}{\sqrt{\tau}}>dt$. We will then consider the limit $\tau\rightarrow0$ in such a way that $G^2\tau$ remains finite. The full dynamics is then:
\begin{align}
    dz_i&=J_i\,dt + \,\sum_j\, G_{ij}\xi_j\,\frac{dt}{\sqrt{\tau}}
   \notag \\
   \text{with}~~~
    J_i &= 4\,J\,z_i\,\left( z_i-\sum_j \,z_j^2\right) \notag \\
    G_{ij}&=2\,\sqrt{\tau G^2}\,z_i\left(\,\delta_{ij}-\,z_j\right) \notag \\
   d\xi_j&=-\xi_j\,\frac{dt}{\tau} + g(\xi_j)dW^j_t
   %\Leftrightarrow\,d\mathbf{z} \,&=\mathbf{J} + \mathbf{G}\mathbf{\xi}
   \label{Eq:ZPair}
\end{align}
Here both indices $i$ and $j$ lie in $[1,N]$, and $G_{ij}$ are components of an $N\times N$ matrix. To extract the limit for $\tau\rightarrow0$, we employ the Kolmogorov Backward equation (KB), which is obtained from the generator, $\Lambda(\mathbf{z},\bm{\xi})$, of the joint stochastic processes $\{z_i,\xi_j\}$. As detailed by Ref.~\cite{SMaritro2}, considering the joint statistics of $\{\mathbf{z},\bm{\xi}\}$ is a necessary step to obtain Markovian dynamics for the coloured noise driven model. This is because the noise is independent of the state dynamics but the state dynamics depends on the noise. Thus, the value $\mathbf{z}_t$ depends on $\bm{\xi}_t=\bm{\xi}_0+\int_0^t\,d\bm{\xi}_t$ which requires information of $\bm{\xi}$ up to time $t$, rendering $\mathbf{z}$ on its own non-Markovian. 
Augmenting the $\mathbf{z}$ with the $\bm{\xi}$ process however, renders the combination Markovian~\cite{SMHanggi94,SMLuczka2005,SMaritro2}. The corresponding KB equation for the joint dynamics is given by:
\begin{align}
    -\partial_t\rho(\mathbf{z},\bm{\xi},t) &= \Lambda(\mathbf{z},\bm{\xi})\,\rho(\mathbf{z},\bm{\xi},t) \\
    \text{where} ~~~~~~
    \Lambda(\mathbf{z},\bm{\xi}) &:= \Lambda_J(\mathbf{z}) + \frac{1}{\sqrt{\tau}}\Lambda_{G}(\mathbf{z},\bm{\xi}) + \frac{1}{\tau}\Lambda_\xi(\bm{\xi}) \\
    \text{and} ~~~~~~
    \Lambda_J(\mathbf{z}) &= \sum_i\,J_i\,\frac{\partial}{\partial\,z_i} \nonumber\\
    \Lambda_G(\mathbf{z},\bm{\xi})  &= \sum_{ij}\xi_j\,G_{ij}\frac{\partial}{\partial\,z_i}\nonumber\\
    \Lambda_\xi(\bm{\xi}) &= \sum_j\,\left(-\xi_j \frac{\partial}{\partial{\xi_j}}+\frac{\tau g^2(\xi_j)}{2}\frac{\partial^2}{\partial\xi^2_j}\right)\label{Eq:Op defs BK}
\end{align}
Here, $\tau g^2(\xi_j)$ is noted to be independent of $\tau$ for both the OU $\left(g(\xi_j):=\sqrt{\frac{2}{\tau}}\right)$ and SBM $\left(g(\xi_j):=\sqrt{\frac{1-\xi_j^2}{\tau}}\right)$ noise processes and further, $i,j\in\,[1,N]$. Note that the operator $\Lambda(\mathbf{z},\bm{\xi},t)$ is a function of variables on the domains of the stochastic processes $z_j(t)$ and $\xi_j(t)$, and in Eq.~\eqref{Eq:Op defs BK} they are interpreted directly as functions of $z_j$ and $\xi_j$ themselves (rather than of the stochastic processes $z_j(t)$ and $\xi_j(t)$). The differential operator $\Lambda_J$ depends on $\mathbf{z}$ alone, while $\Lambda_G$ depends on $\mathbf{z}$ and $\bm{\xi}$, and $\Lambda_\xi$ depends on $\bm{\xi}$ alone, and these dependencies are dropped for ease of notation from here on. This again shows that changes in the quantum state are influenced by the noise, while the noise itself operates independently of the state. 

In the limit of small $\tau$, the coefficient of $\Lambda_\xi$ guarantees that the noise dynamics is always faster than that of the state. Notice however, that we will not assume the noise to always remain in its steady or late-time distribution $\rho^{\infty}(\bm{\xi})$, since assuming so would result in a straightforward, noise-free scenario that fails to account for the time-dependent nature of the dynamics induced by noise in the probability density~\cite{SMLuczka2005,SMHorsthemkeBook2006,SMPavliotis2008}. We will need to assume that such a steady-state distribution exists for the noise, which is true for both the OU ($\rho^{\infty}(\xi)=\frac{1}{\sqrt{2\pi}}e^{\frac{-\xi^2}{2}}$) and SBM ($\rho^{\infty}(\xi)=\frac{1}{2}$) processes. Notice that from the Fokker-Planck-Kolmogorov equations, the steady state distribution $\rho^{\infty}(\bm{\xi}):=\prod_k\rho^\infty(\xi_k)$ (also termed the invariant/null subspace) can be constructed by demanding that the adjoint generator has the action $\Lambda^{*}_\xi\rho^{\infty}(\bm{\xi})=0$.

The Kolmogorov Backward equations of Eq.~\eqref{Eq:Op defs BK} offer a methodical way to construct the probability density $\rho(\mathbf{z},\bm{\xi},t)$ by representing it as a perturbative expansion in $\tau^{1/2}$, treating the correlation time as a small parameter:
\begin{align}
\rho(\mathbf{z},\bm{\xi},t)= \sum^{\infty}_{k=0} \tau^{k/2} \rho_k = \rho_0 + \sqrt{\tau}\rho_1 + \mathcal{O}[\tau]
\label{Eq:rho_tau_exp}
\end{align}
Utilizing Eq.~\eqref{Eq:Op defs BK}, we derive distinct equations for each power of $\tau$:
\begin{align}
    \tau^{-1}:~~~~~~ -\Lambda_{\xi}\rho_0 &=0 \notag\\
    \tau^{-1/2}:~~~~~~ -\Lambda_{\xi}\rho_1& =\Lambda_G\rho_0 \notag \\
    \tau^{0}:~~~~~~ -\Lambda_{\xi}\rho_2&=(\Lambda_J+\partial_t)\rho_0 + \Lambda_G\rho_1 \notag\\
    \tau^{p/2}:~~~ -\Lambda_{\xi}\rho_{p+2} &=(\Lambda_J+\partial_t)\rho_p + \Lambda_G\rho_{p+1}
    \label{Eq:Pert4}
\end{align}
The final line holds for $p>-2$, with the condition that $\rho_{p}=0$ for all ${p<0}$. 

Notice that because $\Lambda_{\xi}$ is a differential operator in $\bm{\xi}$ only, the first line implies that $\rho_0$ depends only on the state vector and time, denoted as $\rho_0 := \rho_0(\mathbf{z},t)$. The subsequent set of equations comprises expressions like $\Lambda_{\xi}\rho_k=f_k(\mathbf{z},\bm{\xi})$. Solutions to these can be constructed using Fredholm's Alternative theorem~\cite{SMPavliotis2008,SMHorsthemkeBook2006,SMBoninTraversaSmallTau}. In a finite-dimensional Hilbert space, this theorem states that an operator equation $A {x_i}= {y_i}$ with operator $A$ and vectors ${x_i}$ and ${y_i}$, has a solution if $n\cdot y_i=0$ for vectors $n$ in the null subspace of the adjoint operator $A^*$ (i.e., $A^{*}{n}=0$). The solvability condition ${n}\cdot {y_i}=0~\forall i$ can be utilized to construct the solution for the system of equations. This is accomplished by repeated operations of $x_i=A^{-1}\,y_i$ which guarantees a solution if $y_i$ is orthogonal to the null subspace.

This method applies to expressions like $\Lambda_{\xi}\rho_k=f_k(\mathbf{z},\bm{\xi})$ in Eqs.~\eqref{Eq:Pert4} because the adjoint operator $\Lambda_{\xi}^*$ describes the forward evolution in Fokker-Planck-Kolmogorov equations of the noise alone~(see previous sections and Refs.~\cite{SMRisken1996,SMgardiner2004handbook,SMHorsthemkeBook2006,SMPavliotis2008}). As both SBM and OU processes allow long-term steady-state probability distributions, we have $\Lambda_{\xi}^{*}\rho^{\infty}(\bm{\xi})=0$, defining the null subspace of the adjoint operator. Further, the orthogonality of a generic function $y(\mathbf{z},\bm{\xi})$ with the null subspace is expressed through the (function space) inner product: $\mathbb{E}^{\infty}_{\xi}[y(\mathbf{z},\bm{\xi})]:=\int d\bm{\xi}\,\rho^\infty(\bm{\xi}) y(\bm{z},\bm{\xi}) =0$. With this, the solvability condition for the $O(\tau^{-1/2})$ equation becomes $\mathbb{E}^{\infty}_{\xi}[\Lambda_G\rho_0]=0$. This condition, termed the centering condition, characterizes the coupling between the state dynamics and the noise~\cite{SMPavliotis2008,SMHorsthemkeBook2006,SMBoninTraversaSmallTau}. For the OU and SBM processes, the ergodicity properties and the linear coupling (a separate $\xi_i$ for each $z_i$), guarantee that the centering condition is satisfied. To see this, note that the form of $\Lambda_G$ has a differential operator only in $\mathbf{z}$ and since  $\mathbb{E}^{\infty}_{\xi}[\xi]=0$, we trivially have $\mathbb{E}^{\infty}_{\xi}[\Lambda_G\rho_0]=0$.  Thus, a solution for the first-order component $\rho_1$ exists, given $\rho_0$, as $\rho_1 = - \Lambda^{-1}_{\xi}\Lambda_G\rho_0$. 

Substituting this $\rho_1$ expression into the equation for $\Lambda_{\xi}\rho_2$ yields the subsequent solvability condition:
\begin{align}
    \mathbb{E}^{\infty}_{\xi}\left[ (\Lambda_J+\partial_t)\rho_0 -\Lambda_G\Lambda^{-1}_{\xi}\Lambda_G\rho_0 \right]=0
    \label{Eq:Final_Solvbl0}
\end{align}
Only the final term on the right-hand side relies on $\xi$, which simplifies this expression to:
\begin{align}
    (\Lambda_J+\partial_t)\rho_0 -\mathbb{E}^{\infty}_{\xi}\left[\Lambda_G\Lambda^{-1}_{\xi}\Lambda_G\rho_0 \right]=0
    \label{Eq:Final_Solvbl}
\end{align}
The remaining expectation value is computed using the so-called cell problem Ansatz~\cite{SMPavliotis2008,SMHorsthemkeBook2006,SMBoninTraversaSmallTau}. This begins with noticing that $\Lambda_G = \sum_j \xi_j \,G_{ij}\frac{\partial}{\partial\,z_j}$, making $\Lambda^{-1}_{\xi}\Lambda_G\rho_0$ expressible as $\Lambda^{-1}_{\xi} (\sum_k\xi_k F_k(\mathbf{z}))$ for an arbitrary function $F_j(\mathbf{z})$. The cell problem Ansatz then posits the existence of a function $\Phi(\mathbf{z},\bm{\xi})$ such that $\Lambda^{-1}_\xi\,\Lambda_G\rho_0 = \Phi(\mathbf{z},\bm{\xi}) $. If this function exists, it implies $\Lambda_{\xi} \Phi(\mathbf{z},\bm{\xi}) = \Lambda_G\rho_0$. Finding such a function $\Phi(\mathbf{z},\bm{\xi})$ is equivalent to determining an expression for $\Lambda^{-1}_{\xi}\Lambda_G\rho_0$ required for evaluating the solvability condition. For our case, using $\Lambda_{\xi}$, the function, $$\Phi(\mathbf{z},\bm{\xi}) = -\sum_{ij}\xi_i\,G_{ij}\frac{\partial}{\partial\,z_j}\rho_0,$$ solves the cell Ansatz for both the OU and SBM noise processes. Now, since, $\Lambda_G\Lambda^{-1}_\xi\,\Lambda_G\rho_0 = \Lambda_G\Phi(\mathbf{z},\bm{\xi}) $, we may evaluate the solvability condition in Eq.~\eqref{Eq:Final_Solvbl} by substituting this into the expectation value, which yields:
\begin{align}
\mathbb{E}^{\infty}_{\xi}\left[\Lambda_G\Lambda^{-1}_{\xi}\Lambda_G\rho_0 \right]=-\mathbb{E}^{\infty}_{\xi}\left[ \xi^2\right]\sum_{i,j,k} \Bigg( G_{ik}\frac{\partial G_{jk}}{\partial\,z_i}\,\frac{\partial\rho_0}{\partial\,z_j} +  G_{ik}\,G_{jk}\frac{\partial^2\rho_0}{\partial\,z_i\partial\,z_j}\,\Bigg)
\label{Eq:Final_Solvbl2}
\end{align}
Here, we used the fact that all $\xi_i$ are identical stochastic processes with the same variance $\mathbb{E}^{\infty}_{\xi}\left[ \xi^2_i\right]=\mathbb{E}^{\infty}_{\xi}\left[ \xi^2\right]$ and $\mathbb{E}^{\infty}_{\xi}\left[ \xi_i\xi_j\right]=0$.

Together with Eq.~\eqref{Eq:Final_Solvbl}, Eq.~\eqref{Eq:Final_Solvbl2} establishes the solvability condition for dynamics up to order $\tau^0$. This homogenizes the stochastic variable to order $\tau^0$, providing an expression for the time evolution of probabilities for $\mathbf{z}$ alone:
\begin{align}
&\left[\partial_t + \Lambda_J +\tilde{\Lambda}_G \right] \rho_0(\mathbf{z},t) =0\notag\\
\tilde{\Lambda}_G&:=\frac{1}{2}\sum_{i,j,k} \Bigg( \tilde{G}_{ik}\frac{\partial \tilde{G}_{jk}}{\partial\,z_i}\,\frac{\partial\rho_0}{\partial\,z_j} +  \tilde{G}_{ik}\,\tilde{G}_{jk}\frac{\partial^2\rho_0}{\partial\,z_i\partial\,z_j}\,\Bigg)
\label{eq:stratBK}
\end{align}
Here, we rescaled $G_{ij}$ by introducing $\tilde{G}_{ij}:=2\sqrt{\mathcal{D}}\,z_i\left(\,\delta_{ij}-\,z_j\right)
$, with the effective coupling defined as $\mathcal{D}=2\mathbb{E}^{\infty}_{\xi}\left[ \xi^2\right]G^2\tau $. This equation signifies the solvability condition for the system of Eq.~\eqref{Eq:ZPair}, and the solutions of Eq.~\eqref{eq:stratBK} weakly correspond (agreeing on ensemble-averaged probabilities $\rho(\mathbf{z},t)$) to those of Eq.~\eqref{Eq:ZPair} in the $\tau\to 0$ limit, when the probability distribution $\rho(\mathbf{z},\bm{\xi},t)$ equals $\rho_0(\mathbf{z},t)$. Furthermore, Eq.~\eqref{eq:stratBK} represents a Kolmogorov Backward equation (in the Stratonovich representation) for the time evolution of the probabilities of $\mathbf{z}$ alone (after homogenizing over the noise)~\cite{SMgardiner2004handbook,SMHanggi94,SMoksendal2003stochastic}. Notably, this equation straightforwardly coincides with the white-noise driven process~\cite{SMPavliotis2008,SMHorsthemkeBook2006,SMBoninTraversaSmallTau, SMgardiner2004handbook,SMRisken1996,SMoksendal2003stochastic}:
\begin{align}
    d\,z_i = J_i\,dt + \sum_j\tilde{G}_{ij}\circ\,dW^j_t\,+\mathcal{O}[\tau]
\end{align}

Because in the Stratonovich representation (where $\circ$ denotes the Stratonovich product) the usual rules of calculus apply, the quantum state reduction dynamics on the Hilbert space for the original pair of processes involving both $\psi$ and $\xi_k$, is now approximated in the $\tau\rightarrow 0$ limit by a single effective stochastic process of the quantum state on the Hilbert space (Eq.~(3) in the main text), given by:
\begin{align}
    d\ket{\psi} = \,\sum_k\,\,  2J\langle\hat{\mathbb{P}}_k\rangle\bigg[\hat{\mathbb{P}}_k -\langle\hat{\mathbb{P}}_k\rangle\bigg]\ket{\psi}dt+\,\sqrt{\mathcal{D}}\,\sum_k\,\bigg[\hat{\mathbb{P}_k} -\langle\hat{\mathbb{P}}_k\rangle\bigg]\ket{\psi}\circ dW_t^k
    \label{Eq:WhiteNoiseSUVFinal}
\end{align}

This establishes the existence of an analytically tractable Markovian limit in the joint state-noise dynamics of Eq.~\eqref{Eq:App_Pair}, such that in the limit of $\tau\rightarrow0$ the dynamics it is weakly equivalent to the white noise driven process of Eq.~\eqref{Eq:WhiteNoiseSUVFinal}. This is an expected result, in light of the Wong-Zakai theorems~\cite{SMWongZakaiReview,SMwongZakai1965convergence,SMWongZakai1969,SMWongZakai1965relation}, and can be generalized further. In the situation of interest in the main text, the Markovian limit (for $\tau\rightarrow0$) is thus found to follow from the straightforward prescription (Eq.~(7) in the main text):
\begin{align}
    \lim_{\tau\small\to0} G \int \left(\hat{\mathbb{P}}_k-\langle\hat{\mathbb{P}}_k\rangle\right) |\psi\rangle \,\xi^k_t \,dt ~~ \to ~~  \sqrt{\mathcal{D}}\,\int \left(\hat{\mathbb{P}}_k-\langle\hat{\mathbb{P}}_k\rangle\right) |\psi\rangle \circ dW^k_t
\end{align}

\section{Emergent Born's rules and GKSL Master equations}
In this section we show the emergence of Born rule statistics in the model defined by Eq.~(3) of the main text. The dynamics follows a Stratonovich stochastic differential equation and this section discusses how to transform to It\^o's convention and then how to average out the stochastic component.
\subsection{The Stratonovich correction}
Following the multi-scale noise homogenization, the dynamics of the quantum state reduction process discussed in the main text (Eq.~(3) of the main text), is given by:
\begin{align}
d|\psi\rangle= 2J\sum^N_{k=0}\,\langle\hat{\mathbb{P}}_k\rangle \left(\hat{\mathbb{P}}_k-\langle\hat{\mathbb{P}}_k\rangle\right)   \ket{\psi}\,dt +\sqrt{\mathcal{D}}\sum^N_{k=0} \left(\hat{\mathbb{P}}_k-\langle\hat{\mathbb{P}}_k\rangle\right)  \ket{\psi}\circ dW^k_t.
\label{Eq:Supp1}
\end{align}
Employing the Stratonovich to It\^o transformation, $X_t\circ dW_t=X_t dW_t+\frac{1}{2}dX_t dW_t$~\cite{SMoksendal2003stochastic} with $X_t=\sqrt{\mathcal{D}}\sum_k\big(\hat{\mathbb{P}}_k-\langle\hat{\mathbb{P}}_k\rangle\big)\ket{\psi}$ results in the It\^o form:
$$
\sqrt{\mathcal{D}}\sum^N_{k=0} \left(\hat{\mathbb{P}}_k-\langle\hat{\mathbb{P}}_k\rangle\right)  \ket{\psi}\circ dW^k_t = \sqrt{\mathcal{D}}\sum^N_{k=0} \left(\hat{\mathbb{P}}_k-\langle\hat{\mathbb{P}}_k\rangle\right)  \ket{\psi} dW^k_t + \hat{C}_t\ket{\psi}dt.
$$
Here, the Stratonovich correction, $\hat{C}_t$, is given by:
$$
\hat{C}_t=\mathcal{D}\sum^{N}_{k=0}\bigg(\frac{1}{2} 
    \left[\hat{\mathbb{P}}_k - \langle\hat{\mathbb{P}}_k\rangle\right]^2 
    -  \left[\langle\hat{\mathbb{P}}_k^2\rangle - \langle\hat{\mathbb{P}}_k\rangle^2\right]\bigg )
$$
The norm-preserving nature of the dynamics may be confirmed by considering the change in the norm, $dN_t$ where $N_t = \langle\psi_t\ket{\psi_t}$. Here we retained the temporal argument for clarity. Using It\^o's lemma  (i.e. the It\^o-Leibnitz rule, $dN_t= \bra{d\psi_t}\psi_t\rangle + \bra{\psi_t}d\psi_t\rangle +\bra{d\psi_t}d\psi_t\rangle \,\,$) we find $dN_t =0$, which shows that a normalized inital state remains normalized.

\subsection{Master equations}
In this subsection we obtain the noise-averaged master equations (Eq.~(5) from Eq.~(3) in the main text). We retain the time arguments in the following subsections for clarity. To find the evolution of the pure state projector, $\hat{\Psi}_t:= \ket{\psi_t}\bra{\psi_t}$, we use It\^o's lemma, $d\hat{\Psi}_t=|d\psi_t\rangle\langle\psi_t|+|\psi_t\rangle\langle d\psi_t|+|d\psi_t\rangle\langle d\psi_t|$, and obtain:
\begin{align}
d\hat{\Psi}_t = 2J\sum^{N}_{k=0}&\bigg\{\langle\hat{\mathbb{P}}_k\rangle \left(\hat{\mathbb{P}}_k-\langle\hat{\mathbb{P}}_k\rangle\right) ,  \hat{\Psi}_t\bigg\}\,dt +\sqrt{\mathcal{D}} \sum^{N}_{k=0} \bigg\{ \hat{\mathbb{P}}_k-\langle\hat{\mathbb{P}}_k\rangle\, ,\, \hat{\Psi}_t\bigg\}\, dW^k_t  \nonumber\\ +& \bigg\{\hat{C}_t\,,\,\hat{\Psi}_t\bigg\}dt + \sum^{N}_{k=0}\mathcal{D} \,  \left(\hat{\mathbb{P}}_k-\langle\hat{\mathbb{P}}_k\rangle\right)\hat{\Psi}_t \left(\hat{\mathbb{P}}_k-\langle\hat{\mathbb{P}}_k\rangle\right)dt
\end{align}
This expression can be simplified to yield:
\begin{align}
d\hat{\Psi}_t = 2(J-\mathcal{D}) \sum_k \bigg[ \bigg\{ \hat{\mathbb{P}}_k \langle\hat{\mathbb{P}}_k\rangle_t \,,\hat{\Psi}\bigg\} - 2 \langle\hat{\mathbb{P}}_k\rangle_t^2 \hat{\Psi}_t
\, \bigg] dt
+ \mathcal{D} \bigg[ \sum_k \hat{\mathbb{P}}_k \hat{\Psi}_t\hat{\mathbb{P}}_k - \hat{\Psi}_t\bigg]dt +\sqrt{\mathcal{D}} \sum^{N}_{k=0} \bigg\{ \hat{\mathbb{P}}_k-\langle\hat{\mathbb{P}}_k\rangle\, ,\, \hat{\Psi}_t\bigg\} dW^k_t \notag
\end{align}
It is straightforward to see that imposing the fluctuation-dissipation relationship (FDR) of $J=\mathcal{D}$ and averaging over the noise (using $\mathbb{E}[\hat{\Psi}_t]=\hat{\rho}_t$ and $\mathbb{E}[M_tdW_t]=0$ in It\^o's interpretation for arbitrary $M_t$), yields a linear master equation corresponding to a Gorini-Kossakowsky-Sudarshan-Lindblad (GKSL) equation for dephasing, given by:
$$\frac{\partial\hat{\rho}_t}{\partial\,t} = 
 \mathcal{D} \bigg[ \sum_k \hat{\mathbb{P}}_k \hat{\rho}_t\hat{\mathbb{P}}_k - \hat{\rho}_t\bigg]$$
This implies the emergence of Born rule statistics and the absence of superluminal signalling. Using the notation in the main text, $J=\mathcal{J}\mathcal{N}/\hbar$, we obtain $\sqrt{\mathcal{D}}=\sqrt{\mathcal{J}\mathcal{N}/\hbar}$ as written in the main article.
 
\section{The continuum limit}
In this section we discuss the continuum limit of Eq.~(3) in the main text and obtain Eq.~(6) of the main text. We first generalize the notion of a Wiener process to higher dimensions, and then discuss the continuum limit of the white-noise model.

\subsection{Space-time Wiener process from space-time white noise}
Although there are many way to characterize the fundamental Wiener process, not all may be generalized to multi-parameter Brownian motion. The standard characterization of the Wiener process as a definite integral of (Gaussian) white noise, written in terms of independent Gaussian increments and (almost surely) continuous paths, can be generalised to multiple dimensions. This yields the so called Hida's calculus, which allows for a rigorous way to study stochastic partial differential equations on infinite dimensional spaces~\cite{SMoksendal2022spacetime,SMHida1980,SMHolden2009,SMhairer_STWN3}. For clarity, we retain the standard physics notation in the following and use a simplified approach to showcase explicit calculations. 

The Wiener processes in space and time may be considered in analogy to the construction of a Wiener process in time ($t\in[0,T]$), using Gaussian white noise $\eta_t$ sampled at each $t$ from a standard normal distribution with zero mean and unit variance. Note that this construction formally assumes a discretization, such that $\eta_t$ is sampled from $N(0,\sigma^2 = 1)$ at each infinitesimal time step. Here, we follow the physics literature and use integral notation, while discrete summations are preferred in the mathematical literature. The Wiener process is defined as the temporal integral (sum) of $\eta_t$, i.e. $W_t = \int^t ds \,\eta_s$. This can be used to define $W_{t+dt}-W_{t}=dW_t = \int^{t+dt}_t \eta_z\, dz$ (usually written simply as $dW_t = \eta_t\, dt$), which implies that $dW_t$ is sampled from a Gaussian distribution, $N(0, \sigma^2=dt)$, with variance $dt$. From, $\mathbb{E}[\eta_t]=0$ it follows that $\mathbb{E}[dW_t]=0$. Further, since $\mathbb{E}[\eta_t \eta_{s}]=\delta(t-s)$, we obtain the shorthand It\^o multiplication rules:
\begin{align}
\mathbb{E}[\eta_t \eta_{s}] &= \delta(t-s)\\
\implies \mathbb{E}[\,\int_T\eta_z\, dz\, \int_S\eta_{\bar{z}}d\bar{z}\,] &= \int_T dz\,\int_S d\bar{z}\,\,\delta(z-\bar{z})\\
\implies \mathbb{E}[\,dW_t\,dW_s\,] &= \int_T dz\,\int_S d\bar{z}\,\,\delta(z-\bar{z})
\end{align}
Here, we denoted the interval $[t,t+dt]$ as $T$ and $[s,s+dt]$ as $S$, and in the value of the double integral depends on the intervals of integration. If $T \cap S=\emptyset$ the double integral vanishes, while if $T\cap S\neq\emptyset$ it must be the case that $t=s$ and $T=S$ because $T$ and $S$ are infinitesimal intervals. In this case the integral on the right hand side yields the interval size, $dt$, and the Ito multiplication rule becomes $\mathbb{E}[dW^2_t]= dt$ and $\mathbb{E}[\,dW_t\,dW_s\,]= 0$ for all $t\neq s$.

Extending these considerations, space-time white (Gaussian) noise may be defined on both space ($x\in \mathbb{R}$) and time ($t\in[0,T]$), and denoted by $\eta^x_t=\eta(x,t)$. In direct analogy to temporal noise, $\eta^x_t=$ is sampled from the standard normal distribution $N(0,\sigma^2=1)$ at each space-time point, which formally is done using a discretization. The space-time generalization of the Wiener process (often called a Brownian sheet) follows by constructing the space-time double integral, i.e. $\int_T dW_t(X) = \int_X \, dx \int_T dt\,\eta^x_t \,$ where $X$ is an interval in $\mathbb{R}$. Using, $\mathbb{E}[\eta^x_t]=0$, and defining $dW_t^x:=dW_t(dx)$, we again find, $\mathbb{E}[dW^x_t]=0$. To obtain the corresponding It\^o multiplication rules, we utilize $\mathbb{E}[\eta^x_t \eta^y_{s}]=\delta(t-s)\delta(x-y)$, change variables for convenience and consider the following integrals:
\begin{align}
\mathbb{E}[\eta^x_t \eta^y_{s}] &= \delta(t-s)\delta(x-y)\\
\implies \mathbb{E}\bigg[\,\bigg(\int_T dz\int_X dx\,\eta^{x}_z\,\bigg) \,\bigg( \int_S d\bar{z} \int_Y \,dy\, \eta^{y}_{\bar{z}}\,\bigg)\bigg] &= \int_T dz\,\int_S d\bar{z}\,\int_X \,dx\,\int_Y \,dy\,\delta(z-\bar{z})\delta(x-y)\\
\implies \mathbb{E}[\,dW_t(X)\,dW_s(Y)\,] &= \int_T dz\,\int_S d\bar{z}\,\int_X \,dx\,\int_Y \,dy\,\delta(z-\bar{z})\delta(x-y)
\end{align}
In the final expression the quadruple integral is non-zero only if both the temporal intervals ($T$ and $S$) and the spatial intervals ($X$ and $Y$) overlap. For infinitesimal intervals $X=[x,x+dx]$ and $Y=[y,y+dx]$, an overlap is possible only if $x=y$ ($X=Y$) and $t=s$ ($T=S$). This implies the It\^o rules $\mathbb{E}[(dW_t^x)^2]=dt\,dx$ and $\mathbb{E}[dW_t^x dW_s^y]=0$ for all $t\neq s$ or $x \neq y$. This implies It\^o's isometry:
\begin{align}
    \mathbb{E}\bigg[\left(\int_X\int_T G(x,t) dW^x_t\right)^2\bigg]= \int_X\int_T G(x,t) ^2 dtdx
\end{align}
There is no expectation on the right hand side, since the functional form of $G(x,t)$ does not depend on $W^x_t$ explicitly. The above construction is a simple approach to space-time white noise and its connection to the Brownian sheet construction for $X\in\mathbb{R}$, following Refs.~\cite{SMHolden2009,SMoksendal2022spacetime,SMbrehier2014,SMLanconelli2007_STWN0}, which suffices for our needs. More rigorous constructions may be found throughout the mathematical literature. In particular cases where the set $X$ is $\mathbb{R}^n$ instead of just $\mathbb{R}$ as considered here, further spatial noise correlations leading to $\mathbb{E}[dW^x_t\,dW_t^y]=c(x-y,t)dtdx$ need to be taken into account, as discussed for example in Refs.~\cite{SMdalang2010_STWN,SMSTWN2,SMhairer_STWN3,SMHida1980,SMHolden2009}. Such correlations may play a role in more general models of quantum state reduction, which avoid collapse into Dirac-delta localized states in the continuum. We leave this as a possible avenue for further research. From this point on, we adopt the notation $(dW_t^x)^2=dtdx$, while all other squared infinitesimals are zero.

\subsection{Continuum limit of the white-noise model}
Here we discuss the continuum limit of the white noise driven model defined by Eq.~(3) in the main text and obtain the continuum master equation in Eq.~(6). Recall that in previous sections, we showed that a wave function of the form $\ket{\psi_t}=\sum^N_{k=1}\,\psi_k(t)\,\ket{k}$ collapses to a single basis-vector in $\{\ket{k}\}$ under the dynamics defined by:
\begin{align}
    d\ket{\psi_t} &=  \hat{J}_t\ket{\psi_t}dt \,+\hat{C}_t\ket{\psi_t}dt +d\hat{G}_t\ket{\psi_t}
\end{align}    
Here, we used the operators defined by:    
\begin{align}
    \hat{J}_t\ket{\psi_t}dt&:=2J\sum_k \,\bigg(\hat{\mathbb{P}}_k\langle\hat{\mathbb{P}}_k\rangle-\langle\hat{\mathbb{P}}_k\rangle^2\bigg)\ket{\psi_t}dt\nonumber\\
    \hat{C}_t\ket{\psi_t}dt&:=\mathcal{D}\sum_{k}\bigg(\frac{1}{2} 
    \left[\hat{\mathbb{P}}_k - \langle\hat{\mathbb{P}}_k\rangle\right]^2 
    -  \left[\langle\hat{\mathbb{P}}_k^2\rangle - \langle\hat{\mathbb{P}}_k\rangle^2\right]\bigg )\ket{\psi_t}dt\nonumber\\
    d\hat{G}_t\ket{\psi_t}&:=\sqrt{\mathcal{D}}\sum_{k} \left(\hat{\mathbb{P}}_k-\langle\hat{\mathbb{P}}_k\rangle\right) \ket{\psi_t} dW^k_t
\end{align}    
These can be written out in terms of wave function components as:    
\begin{align}    
    \hat{J}_t\ket{\psi_t}dt&=2J\left[\left(\sum_k |\psi^k_t|^2\,\hat{\mathbb{P}}_k\right)-\left(\sum_k|\psi^j_t|^4\right)\hat{\mathbb{1}}\right]\ket{\psi_t}dt
    \nonumber\\
    \hat{C}_t\ket{\psi_t}dt&=\mathcal{D}\bigg[\left(\frac{3}{2}\sum_k|\psi^k_t|^4 - \frac{1}{2}\right)\hat{\mathbb{1}} -\left(  \sum_{k}|\psi^k_t|^2\,\hat{\mathbb{P}}_k\right)\bigg]\ket{\psi_t}dt
    \nonumber\\
    d\hat{G}_t\ket{\psi_t}&=\sqrt{\mathcal{D}}\sum_{k} \left(\hat{\mathbb{P}}_k-\langle\hat{\mathbb{P}}_k\rangle\right) \ket{\psi_t} dW^k_t
\end{align}

The continuum limit equations for a wave-function defined on the real line, i.e. $\ket{\psi}_t=\int_{X=\mathbb{R}} \psi_t(x)\ket{x}dx$, may now be straightforwardly obtained by using the notion of Brownian sheet discussed in the last section along with its generalized It\^o's multiplication rules. Using the projector valued measure $d\hat{\mathbb{P}}_x:=\ket{x}\bra{x}\,dx$ and the projector valued stochastic integral $\hat{\mathbb{P}}_x\,dW^x_t = \ket{x}\bra{x}\,dW_t^x$, we have the following continuum limit identifications:
\begin{align}
    2J\left[\left(\sum_k |\psi^k_t|^2\,\hat{\mathbb{P}}_k\right)-\left(\sum_k|\psi^j_t|^4\right)\hat{\mathbb{1}}\right]\ket{\psi_t}dt&\rightarrow\,2J\left[\left(\int_X |\psi_t(x)|^2\,d\hat{\mathbb{P}}_x\right)-\left(\int_X\,dx\,|\psi_t(x)|^4\right)\hat{\mathbb{1}}\right]\ket{\psi_t}dt\nonumber\\
    \mathcal{D}\bigg[\left(\frac{3}{2}\sum_k|\psi^k_t|^4 - \frac{1}{2}\right)\hat{\mathbb{1}} -\left(  \sum_{k}|\psi^k_t|^2\,\hat{\mathbb{P}}_k\right)\bigg]\ket{\psi_t}dt    &\rightarrow  \mathcal{D}\bigg[\left(\frac{3}{2}\int_X\,dx\,|\psi_t(x)|^4 - \frac{1}{2}\right)\hat{\mathbb{1}} -\left(  \int_X|\psi_t(x)|^2\,d\hat{\mathbb{P}}_x\right)\bigg]\ket{\psi_t}dt   \nonumber\\
    \sqrt{\mathcal{D}}\sum_{k} \left(\hat{\mathbb{P}}_k-\langle\hat{\mathbb{P}}_k\rangle\right) \ket{\psi_t} dW^k_t&\rightarrow \sqrt{\mathcal{D}}\int_X \left(\hat{\mathbb{P}}_x-|\psi_t(x)|^2\,\hat{\mathbb{1}}\right) \ket{\psi_t} dW^x_t
\end{align}
Note that both integrals in time and space are now explicit and the modified Schr\"odinger equation reads $d\ket{\psi_t} =  \hat{J}_t\ket{\psi_t}dt \,+\hat{C}_t\ket{\psi_t}dt +d\hat{G}_t\ket{\psi_t}$ with:
\begin{align}
   \hat{J}_t\ket{\psi_t}dt&:=2\,J\left[\left(\int_X |\psi_t(x)|^2\,d\hat{\mathbb{P}}_x\right)-\left(\int_X\,dx\,|\psi_t(x)|^4\right)\hat{\mathbb{1}}\right]\ket{\psi_t}dt\\
    \hat{C}_t\ket{\psi_t}dt&:=  \mathcal{D}\bigg[\left(\frac{3}{2}\int_X\,dx\,|\psi_t(x)|^4 - \frac{1}{2}\right)\hat{\mathbb{1}} -\left(  \int_X|\psi_t(x)|^2\,d\hat{\mathbb{P}}_x\right)\bigg]\ket{\psi_t}dt   \\
    d\hat{G}_t\ket{\psi_t}&:=\sqrt{\mathcal{D}}\int_X \left(\hat{\mathbb{P}}_x-|\psi_t(x)|^2\,\hat{\mathbb{1}}\right) \ket{\psi_t} dW^x_t
\end{align}
In the next section, the fluctuation dissipation relation $J=\mathcal{D}$ is shown to yield emergent Born rule statistics as well as a linear master equation. Using this relation, we obtain the state evolution equations:
\begin{align}
   d\ket{\psi}=\,J\left[\left(\int_X |\psi_t(x)|^2\,d\hat{\mathbb{P}}_x\right)-\frac{1}{2}\left(1+\int_X\,dx\,|\psi_t(x)|^4\right)\hat{\mathbb{1}}\right]\ket{\psi_t}dt+\sqrt{J}\int_X \left(\hat{\mathbb{P}}_x-|\psi_t(x)|^2\,\hat{\mathbb{1}}\right) \ket{\psi_t} dW^x_t
\end{align}
\subsubsection{Normalization}
To check for norm preservation we compute the evolution of $N_t = \bra{\psi_t}\psi_t\rangle$ with It\^o's lemma (the It\^o-Leibnitz rule), $dN_t= \bra{d\psi_t}\psi_t\rangle + \bra{\psi_t}d\psi_t\rangle +\bra{d\psi_t}d\psi_t\rangle $:
\begin{align}
    dN_t &= 2\langle\hat{J}_t\rangle\,dt + 2\langle\hat{C}_t\rangle\,dt + 2\langle\,d\hat{G}_t\rangle+ \langle d\hat{G}^2_t\rangle \notag \\
    &=2\langle\hat{C_t}\rangle\,dt +  \langle d\hat{G}^2_t\rangle=0
\end{align}
Here, $\langle\hat{O}\rangle=\bra{\psi_t}\hat{O}\ket{\psi_t}/\langle\psi_t\ket{\psi_t}$ is the usual quantum expectation value and the second equality follows from $\langle\hat{J}_t\rangle_t = \langle\,d\hat{G}\rangle_t=0$. The Stratonovich correction term can be written as $2\langle \,\hat{C}_t\rangle\,dt= \left(\int_X\, dx\,|\psi_t(x)|^4 -1\right)dt$, while for It\^o's correction term we use $(dW^x_t)^2=dt\,dx$ (and zero otherwise) to find $\langle d\hat{G}^2_t\rangle=-\left(\int_X\, dx\,|\psi_t(x)|^4 -1\right)dt$. Combining these yields $dN_t=0$, indicating that the norm is preserved by the dynamics.

\subsection{Continuum Master Equations}
We obtain the continuum dynamical master equations by computing the evolution of the pure state projection operator $\hat{\Psi}_t=\ket{\psi_t}\bra{\psi_t}$ using It\^o's lemma ($d\hat{\Psi}_t=|d\psi_t\rangle\langle\psi_t|+|\psi_t\rangle\langle d\psi_t|+|d\psi_t\rangle\langle d\psi_t|$) and averaging out the noise. For the pure state projector,this yields:
\begin{align}
    d\hat{\Psi}_t \,=\, \left\{\hat{J}_t\,,\,\hat{\Psi}_t\right\}dt + \left\{\hat{C}_t\,,\,\hat{\Psi}_t\right\}dt + \left\{d\hat{G}_t\,,\,\hat{\Psi}_t\right\} + d\hat{G}_t\,\hat{\Psi}_t\,d\hat{G}_t 
\end{align}
Term by term, this can be written as:
\begin{align}
    \left\{\hat{J}_t +\hat{C}_t\, ,\,\hat{\Psi}\right\} dt &=  \, \left\{ \big(2J-\mathcal{D}\big)\bigg(\int_x\,|\psi_t(x)|^2\,d\hat{\mathbb{P}}_x\bigg)\,-\big(2J-\frac{3}{2}\mathcal{D}\big)\bigg(\int_X\,dx\,|\psi_t(x)|^4\,\bigg)\hat{\mathbb{1}}-\frac{\mathcal{D}}{2}\hat{\mathbb{1}}\,,\hat{\mathbb{\Psi}}_t\,\right\}dt\\
    \left\{d\hat{G}_t\,,\,\hat{\Psi}_t\right\}&= \sqrt{\mathcal{D}}\int_X\left\{ \left(\hat{\mathbb{P}}_x-|\psi_t(x)|^2\,\hat{\mathbb{1}}\right)  \,,\,\hat{\Psi}_t\right\} dW^x_t
\end{align}
The final term corresponds to It\^o's correction, and can be written as:
\begin{align}
    d\hat{G}_t\,\hat{\Psi}_t\,d\hat{G}_t&=\mathcal{D}\int_X\int_Y\,dW^x_t\,dW^y_t\bigg(\hat{\mathbb{P}}_x\hat{\Psi}_t\hat{\mathbb{P}}_y\,+ |\psi_t(x)|^2\,|\psi_t(y)|^2\,\hat{\Psi}_t\, -|\psi_t(y)|^2 \hat{\mathbb{P}}_x\hat{\Psi}_t \,-|\psi_t(y)|^2 \hat{\Psi}_t\hat{\mathbb{P}}_y \bigg)\\
    &=\mathcal{D}\int_X\,dx\bigg[\hat{\mathbb{P}}_x\hat{\Psi}_t\hat{\mathbb{P}}_x\,+ |\psi_t(x)|^4\,\hat{\Psi}_t\, -\left\{|\psi_t(x)|^2 \hat{\mathbb{P}}_x\,,\,\hat{\Psi}_t\right\}  \bigg]dt\\
    &=\mathcal{D}\bigg[\int_X\,d\hat{\mathbb{P}}_x\,\langle\,x\,|\hat{\Psi}_t|\,x\rangle\,+ \frac{1}{2}\,\left\{\left(\int_X\,dx\,|\psi_t(x)|^4\right)\hat{\mathbb{1}}\,,\,\hat{\Psi}_t\right\}\, -\left\{\int_X|\psi_t(x)|^2 d\hat{\mathbb{P}}_x\,,\,\hat{\Psi}_t\right\}  \bigg]dt
\end{align}
In the above, obtaining the second expression from the first utilizes $dW^x_t\,dW^y_t=0$ for all $x\neq y$ and $(dW_t^x)^2=dt\,dx$, as explained in the previous sections. In the last step we rearrange terms and used the projector valued measure $d\hat{\mathbb{P}}_x := \ket{x}\bra{x}dx$, used commonly in the spectral decomposition of continuous unbounded operators. Combining all terms results in the expression:
\begin{align}
    d\hat{\Psi}_t \,= \mathcal{D}\left[\int_X\,d\hat{\mathbb{P}}_x\,\langle\,x\,|\hat{\Psi}_t|\,x\rangle\,-\,\hat{\Psi}_t\right]&dt \,+\, 2\left(J-\mathcal{D}\right)\,\left\{\int_x\,|\psi_t(x)|^2\,d\hat{\mathbb{P}}_x\,-\hat{\mathbb{1}}\bigg(\int_X\,dx\,|\psi_t(x)|^4\,\bigg)\,\,,\hat{\Psi}_t\,\right\}dt\nonumber\\& \, \,+\,\sqrt{\mathcal{D}}\int_X\left\{ \left(\hat{\mathbb{P}}_x-|\psi_t(x)|^2\,\hat{\mathbb{1}}\right)  \,,\,\hat{\Psi}_t\right\} dW^x_t
\end{align}
Using the Fluctuation-dissipation relationship $J=\mathcal{D}$, and averaging out the noise to obtain the density (statistical) operator, i.e. $\mathbb{E}[\hat{\Psi}_t]=\hat{\rho}_t$, we obtain a continuum version of the dephasing GKSL master equation:
\begin{align}
    \frac{\partial\hat{\rho}_t}{\partial\,t}=\mathcal{D}\left[\int_X\,d\hat{\mathbb{P}}_x\,\langle\,x\,|\hat{\rho}_t|\,x\rangle\,-\,\hat{\rho}_t\right]
\end{align}

\end{document}